\documentclass[extra,fleqn]{gji} 

\usepackage{graphicx} 
\usepackage{rotating}

\usepackage{booktabs}
\usepackage{times}
\usepackage{xfrac,stmaryrd}

\usepackage{amssymb}

\usepackage{amsmath,wasysym,textcomp}

\usepackage{amsxtra}
\usepackage{mathcomp}

\usepackage{multirow}

\title{Thermal convection in  Earth's inner core with phase change at its boundary}
\author[R. Deguen , T. Alboussi\`ere and P. Cardin]{Renaud Deguen$^{1}$, Thierry Alboussi\`ere$^2$ and Philippe Cardin$^{3}$ \\
$^{1}$Department of Earth and Planetary Sciences, Johns Hopkins University, Baltimore, MD 21218, USA. \\
$^2$ LGL, Laboratoire de G\'eologie de Lyon, CNRS, Universit\'e Lyon 1, ENS-Lyon, G\'eode,\\ 2 rue Rapha\"el Dubois, 69622 Villeurbanne, France.\\
$^3$ ISTerre, Universit\'e de Grenoble 1, CNRS, B.P. 53, {38041} Grenoble, France.
}

\begin{document}

\maketitle

\begin{abstract}

Inner core translation, with solidification on one hemisphere and melting on the other, provides a promising basis for understanding the hemispherical dichotomy of the inner core, as well as the anomalous stable layer observed at the base of the outer core - the so-called F-layer -  which might be sustained by continuous melting of inner core material.
In this paper, we study in details the dynamics of inner core thermal convection when dynamically induced melting and freezing of the inner core boundary (ICB) are taken into account.

If the inner core is unstably stratified, linear stability analysis and  numerical simulations consistently show that the translation mode dominates only if the viscosity $\eta$ is  large enough, with a critical viscosity value, of order $\sim 3\ 10^{18}$ Pa~s, depending on the ability of outer core convection to supply or remove the latent heat of melting or solidification.
If $\eta$ is smaller, the dynamical effect of melting and freezing is small. 
Convection takes a more classical form, with a one-cell axisymmetric mode at the onset and chaotic plume convection at large Rayleigh number.
$\eta$ being poorly known, either mode seems equally possible.
We derive analytical expressions for the rates of translation and melting for the translation mode, and  a scaling theory for high Rayleigh number plume convection.
Coupling our dynamical models  with a model of inner core thermal evolution, we predict the convection mode and melting rate as functions of inner core age, thermal conductivity, and viscosity.
If the inner core is indeed in the translation regime, the predicted melting rate  is high enough, according to \cite{Alboussiere2010}'s experiments,  to allow the formation of a stratified layer above the ICB.
In the plume convection regime, the melting rate, although smaller than in the translation regime, can still be significant if $\eta$ is not too small.

Thermal convection requires that a superadiabatic temperature profile is maintained in the inner core, 
which depends on a competition between  extraction of the inner core internal heat by conduction and cooling at the ICB.
Inner core thermal convection appears very likely with the low thermal conductivity value proposed by \cite{Stacey2007}, but nearly impossible with the much higher  thermal conductivity recently put forward by \cite{sha2011}, \cite{deKoker2012} and \cite{pozzo2012}.
We argue however that the formation of an iron-rich layer above the ICB may have a positive feedback on inner core convection~: it implies that the inner core crystallized from an increasingly iron-rich liquid, resulting in an unstable compositional stratification which could drive inner core convection, perhaps even if the inner core is subadiabatic.

\end{abstract}


\section{Introduction}
\label{intro}

In the classical model of convection and dynamo action in Earth's outer core, convection is thought to be driven by a combination of cooling from the core-mantle boundary (CMB) and light elements (O, Si, S, ...) and latent heat release at the inner core boundary (ICB). 
Convection is expected to be vigorous, and the core must therefore be  very close to adiabatic, with only minute lateral temperature variations \citep{stevenson87}, except in  very thin, unstable boundary layers at the ICB and CMB. 
To a large extent, seismological models are consistent with the bulk of the core being well-mixed and adiabatic, which  supports the standard model of outer core convection.
Yet seismological observations indicate the existence of significant deviations from adiabaticity in the lowermost $\sim 200$ km of the outer core \citep{Souriau91}. 
This layer, sometimes called F-layer for historical reasons, exhibits an anomalously low $V_P$ gradient which is most probably indicative of stable compositional stratification \citep{Gubbins08}, implying that the lowermost 200 km of the outer core are depleted in light elements  compared to the bulk of the core.
This is in stark contrast with the classical model of outer core convection sketched above: in place of the expected thin unstable boundary layer, seismological models argues for a very thick and stable layer.
Note also that the thickness of the layer, $\sim 200$ km, is much larger than any diffusion length scales, even on a Gy timescale, which means that if real this layer must have been created, and be sustained, by a mechanism involving advective transport.

Because light elements are partitioned  preferentially into the liquid during solidification, iron-rich melt can be produced through a two-stage purification process involving solidification followed by melting \citep{Gubbins08}.
Based on this idea, \cite{Gubbins08} have proposed a model for the formation of the F-layer in which iron-rich crystals nucleate at the top of the layer and melt back as they sink toward the ICB, thus implying a net inward transport of iron which results in a stable stratification.
In contrast, \cite{Alboussiere2010} proposed that melting occurs directly at the ICB in response to inner core internal dynamics, in spite of the fact that the inner core must be crystallizing on average.
Assuming that the inner core is melting in some regions while it is crystallizing in others, the conceptual model proposed by \cite{Alboussiere2010} works as follow : 
melting inner core material produces a dense iron-rich liquid which spreads at the surface of the inner core, while crystallization produces a buoyant liquid which  mixes with and carries along part of the dense melt as it rises. 
The stratified layer results from a dynamic equilibrium between production of iron-rich melt and entrainment and mixing associated with the release of buoyant liquid. 
Analogue fluid dynamics experiments demonstrate the viability of the mechanism, and show that a stratified layer indeed develops if the buoyancy flux associated with the dense melt is larger (in magnitude) than a critical fraction ($\simeq80$ \%) of the buoyancy flux associated with the light liquid.
This number is not definitive because possibly important factors were absent in \cite{Alboussiere2010}'s experiments (Coriolis and Lorentz force, entrainment by thermal convection from above, ...) but it seems likely that a high rate of melt production will still be required.

A plausible way to melt the inner core is to sustain dynamically a topography that will bring locally the ICB at a potential temperature lower than that of the adjacent liquid core, which allows heat to flow from the outer core to the inner core.
The melting rate is then limited by the ability of outer core convection to provide the latent heat absorbed by melting, and only a significant ICB topography can lead to a non-negligible melting rate.
More recently, \cite{Gubbins2011} and \cite{sreenivasan2011} have proposed that localized melting of the inner core might be induced by outer core convection, but the predicted rate of melt production is too small to produce a stratified layer according to \cite{Alboussiere2010}'s experiments. Furthermore, it is not clear that the behavior observed in numerical simulations at slightly supercritical conditions would persist at Earth's core conditions.

\begin{figure}
\begin{center}
\includegraphics[width=0.9\linewidth]{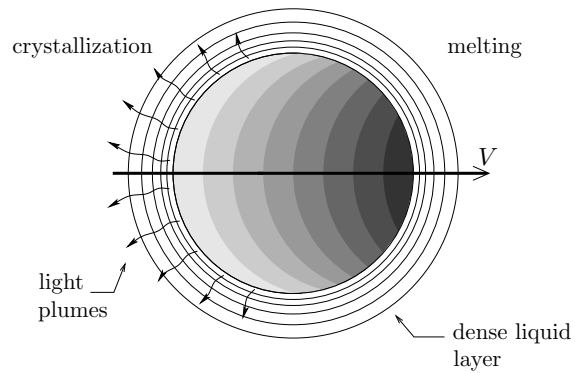}
\end{center}
\caption{A schematic representation of the translation mode of the inner core, with the grey shading showing the potential temperature distribution (or equivalently the density perturbation) in a cross-section including the translation direction (adapted from \citet{Alboussiere2010}). 
\label{schematic}
}
\end{figure}

Among the different models of inner core dynamics proposed so far \citep{Jeanloz1988,Yoshida1996,Karato1999,Buffett2001,Deguen2011b}, only thermal convection \citep{Jeanloz1988,Weber92,Buffett2009,Deguen2011a,Cottaar2012} is potentially able to produce a large dynamic topography and associated melting.
Thermal convection in the inner core is possible if the growth rate of the inner core is large enough and its thermal conductivity low enough \citep{Sumita1995,Buffett2009,Deguen2011a}.
One possible mode of inner core thermal convection consists in  a global translation with solidification on one hemisphere and melting on the other \citep{Monnereau2010,Alboussiere2010,Mizzon2013}.
The translation rate can be such that the rate of melt production is high enough to explain the formation of the F-layer \citep{Alboussiere2010}.
In addition, inner core translation provides a promising basis for understanding the hemispherical dichotomy of the inner core observed in its seismological properties \citep{Tanaka97,Niu01,irving2009,Tanaka2012}.
Textural change of the iron aggregate during the translation \citep{Monnereau2010,bergman2010,Geballe2013} may explain the hemispherical structure of the inner core.
Inner core translation, by imposing a highly asymmetric buoyancy flux at the base of the outer core, is also a promising candidate  \citep{Davies2013,Aubert2013} for explaining the existence of the planetary scale eccentric gyre  which has been inferred from quasi-geostrophic core flow inversions \citep{pais2008,Gillet09}.

However, inner core translation induces horizontal temperature gradients (see Figure \ref{schematic}), and \cite{Alboussiere2010} noted  that finite deformation associated with these density gradients is expected to weaken the translation mode if the inner core viscosity is too small. They estimated from an order of magnitude analysis that the threshold would be at  $\eta \sim 10^{18}$ Pa~s.
Below this threshold,  thermal convection is expected to take a more classical form, with cold plumes falling down from the ICB and warmer upwellings  \citep{Deguen2011a}.
Published estimates of inner core viscosity range from $\sim 10^{11}$ Pa~s to $\sim 10^{22}$ Pa~s \citep{Yoshida1996,Buffett1997,VanOrman04,Koot2011,Reaman2011,Reaman2012} implying that both convection regime seem possible.

The purpose of this paper is twofold : (i) to precise under what conditions the translation mode can be active, and (ii) to estimate the rate of melt production associated with convection, 
in particular when the effect of finite viscosity becomes important.
To this aim, we develop a set of equations for  thermal convection in the inner core with phase change associated with a dynamically sustained topography at the inner core boundary (section \ref{gov_eqns}).
The kinetics of phase change is described by a non-dimensional number, noted $\mathcal{P}$ for "phase change number", which is the ratio of a phase change timescale (introduced in section \ref{phase_change}) to a viscous relaxation timescale. 
The linear stability analysis of the set of equations (section \ref{linear_stability}) demonstrates that the first unstable mode of thermal convection consists in a global translation when $\mathcal{P}$
 is small. When $\mathcal{P}$ is large, the first unstable mode is the classical one cell convective mode of thermal convection in a sphere with an impermeable boundary \citep{Chandrasekhar1961}.
An analytical expression for the rate of translation is derived in section \ref{Small_P}.
We then describe numerical simulation of thermal convection, from which we derive scaling laws for the rate of melt production (section \ref{Numerical_results}). 
The results of the previous sections are then applied to the inner core, and used to predict the convection regime of the inner core and the rate of melt production as functions of the inner core growth rate and thermo-physical parameters  (section \ref{application}).

\section{Phase change at the ICB}
\label{phase_change}

\begin{figure}
\includegraphics[width=0.9\columnwidth]{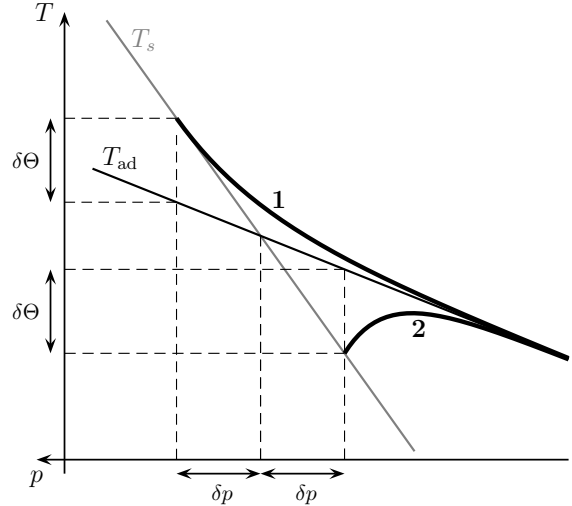}
\caption{Temperature profiles (thick black lines) in the vicinity of the inner core boundary.
Profile \textbf{1} corresponds to a crystallizing region, while profile  \textbf{2} corresponds to a melting region.
The thin black line is the outer core adiabat $T_\mathrm{ad}$ and the thin grey line is the solidification temperature profile.
\label{ICBThermalConditions}
}
\end{figure}

Any phase change at the ICB will release or absorb latent heat, with the rate of phase change $v$ being determined by the Stefan condition, 
\begin{equation}
\rho_s L v = - \llbracket q \rrbracket_\mathrm{icb},  \label{StefanCondition}
\end{equation}
which equates the rate of latent heat release or absorption associated with solidification or melting with the difference of heat flux $\llbracket q\rrbracket_\mathrm{icb}$ across the inner core boundary.
Here $\rho_s$ is the density of the solid inner core just below the ICB, and $L$ is the latent heat of melting. 
The heat conducted along the adiabatic gradient on the outer core side is to a large extent balanced  by the heat flow conducted on the inner core side, the difference between the two making a very small contribution to $\llbracket q\rrbracket_\mathrm{icb}$.
Convective heat transport in the inner core is small as well.
Convection in the liquid outer core is a much more efficient way of providing or removing latent heat and  $- \llbracket q\rrbracket_\mathrm{icb}$ is dominated by the contribution of the advective heat flux $\Phi(\theta,\phi)$ on the liquid side, which scales as
\begin{equation}
\Phi \sim \rho_l c_{pl} u' \delta \Theta, \label{heat}
\end{equation}
where $\delta \Theta$ is the difference of potential temperature between the inner core boundary and the bulk of the core (Figure \ref{ICBThermalConditions}), $u'$ is a typical velocity scale in the outer core, and $\rho_l$  and $c_{pl}$ are the density and specific heat capacity of the liquid outer core in the vicinity of the inner core boundary \citep{Alboussiere2010}.

We choose as a reference radius the intersection of the mean outer core adiabat with the solidification temperature curve (Figure \ref{ICBThermalConditions}), and note $h(\theta,\phi)$ the distance from  this reference to the inner core boundary.
At a given location on the ICB, the difference of potential temperature between the ICB and the outer core is $\delta \Theta(\theta,\phi) = (m_p-m_\mathrm{ad}) \delta p(\theta,\phi)$, where $\delta p(\theta,\phi)$ is the pressure difference between the ICB and the reference surface (see Figure \ref{ICBThermalConditions}), $m_p=dT_s/dp$ is the Clapeyron slope, and $m_\mathrm{ad}=dT_\mathrm{ad}/dp$ is the adiabatic gradient in the outer core.
Taking into account the local anomaly $\Psi '$ of the gravitational potential (due to the ICB topography and internal density perturbations), we have from hydrostatic equilibrium $\delta p =  - \rho_l (g_\mathrm{icb} h + \Psi')$, which gives 
\begin{equation}
\delta \Theta = - (m_p - m_\mathrm{ad}) \rho _l g_\mathrm{icb} \left( h + \frac{\Psi ' }{g_\mathrm{icb}} \right), \label{deltaT}
\end{equation}
where $g_\mathrm{icb}$ is the average gravity level on the surface of the inner core. 
The surface $h_\mathrm{eq}(\theta,\phi)= -{\Psi ' }/{g_\mathrm{icb}}$ is the equipotential surface which  on average coincides with the ICB. 

If the inner core is convecting, with a velocity field ${\bf u}(r,\theta,\phi,t)=(u_r,u_\theta,u_\phi)$, then the total rate of phase change is 
\begin{equation}
v = \dot r_\mathrm{ic} +\frac{ \partial h }{\partial t} - u_r ,   \label{PhaseChangeRate}
\end{equation}
where  $\dot r_\mathrm{ic}$ is the mean inner core growth rate, $r_\mathrm{ic}(t)$ being the inner core radius. 
Using Eqs. \eqref{heat}, \eqref{deltaT} and \eqref{PhaseChangeRate}, the heat balance \eqref{StefanCondition} at the inner core boundary can be written as
\begin{equation}
u_r - \dot r_\mathrm{ic} - \frac{\partial h }{\partial t} \sim \frac{ h + {\Psi ' }/{g_\mathrm{icb}}}{\tau _\phi }, \label{CLheat} \\
\end{equation}
where the time scale $\tau _\phi$ is
\begin{equation}
\tau _\phi = \frac{\rho_s\, L}{\rho _l^2 c_{pl} \left( m_p - m_\mathrm{ad} \right) g_\mathrm{icb} u' }. \label{tauphi} 
\end{equation}
With $u' \sim 10^{-4}$m~s$^{-1}$ and typical values for the other parameters (Table \ref{parameters}), the phase change timescale $\tau _\phi$ is found to be of the order of $10^3$~years, which will turn out to be short compared to the dynamical time-scale of thermal convection in the inner core ($\sim 1$~My or more).
Noting $\Delta \rho=\rho_s-\rho_l$, the viscous relaxation timescale $\tau_\eta = \eta/(\Delta \rho\,g_\mathrm{icb}\,r_\mathrm{ic})$ is at most $\sim 0.1$ My (for $\eta=10^{22}$ Pa~s),  small as well compared to the inner core dynamical timescale. 
We therefore adopt the hypothesis of isostasy and neglect $\partial h / \partial t$ in (\ref{CLheat}), the heat transfer boundary condition finally adopted being written 
\begin{equation}
u_r - \dot r_\mathrm{ic} = \frac{ h + {\Psi ' }/{g_\mathrm{icb}}}{\tau _\phi }, \label{CLheat2} \\
\end{equation}
where the unknown proportionality constant in Equation \eqref{heat} has been absorbed in $\tau_\phi$, and will be treated as an additional source of uncertainty.

\begin{table*}
\caption{Thermo-physical parameters used in this study. \label{parameters}}
{\small
\begin{center}
\begin{tabular}{@{}lll@{}}
\toprule
Parameter 											& Symbol 						& Value 							\\
\midrule
Inner core radius$^a$  								& $r_{\mathrm{icb}}$ 	& $1221$  km				\\ 
Solidification temperature $^b$					& $T_{\mathrm{icb}}$ 		& $5600\pm500$ K		\\ 
Gruneisen parameter $^c$ & $\gamma$					& $1.4\pm0.1$ 			\\
Thermal expansion 	$^c$						& $\alpha$					& $(1.1\pm0.1)\times10^{-5}$	K$^{-1}$			\\ 
Heat capacity		$^d$					 			& $c_p$ 						    & $800\pm80$				  		J~kg$^{-1}$~K$^{-1}$\\ 
Latent heat of melting  $^{d,e}$					& $L$		& $600-1200$ kJ~kg$^{-1}$   \\
Density jump at the ICB $^a$					& $\Delta \rho$							& 600 kg~m$^{-3}$			\\
Density in the inner core	$^a$					& $\rho_s$							& $12\,800$				kg~m$^{-3}$			\\ 
Density in the outer core at the ICB	$^a$					& $\rho_l$							& $12\,200$				kg~m$^{-3}$	\\ 
Gravity at the ICB 	$^a$		& $g_\mathrm{icb}$		&  4.4~m~s$^{-2}$ \\		
Radial gravity gradient	$^a$					& $g'$							& $3.6\ 10^{-6}$ s$^{-2}$  \\
Thermal conductivity  	$^f$						& $k$ 								& $36-150$ 						W$\,$m$^{-1}\,$K$^{-1}$ 		\\
Isentropic bulk modulus	$^a$						& $K_S$							&  $1400$ GPa         \\
Clapeyron/adiabat slopes ratio	$^g$						& $dT_s/dT_{ad}$ & $1.65\pm0.11$  \\
\bottomrule
\end{tabular}
\end{center}
 $^a$ From PREM \citep{PREM}.\\
 $^b$  \cite{Alfe2002}.\\
 $^c$  \cite{Vocadlo2007}.\\
 $^d$  \cite{Poirier1994a} \\ 
 $^e$ \cite{Anderson1997} \\
 $^f$  \cite{Stacey2001}; \cite{Stacey08};  \cite{sha2011}; \cite{deKoker2012}; \cite{pozzo2012}.\\
 $^g$ \cite{Deguen2011a}
  }
\end{table*}

\section{Governing equations}
\label{gov_eqns}

\subsection{Equations within the inner core}

The starting point for the dynamics of thermal convection in the inner core is expressed as general entropy, momentum, continuity and gravitational equations:
\begin{align}
\rho T \frac{D s}{D t} &= {\mitbf{\nabla}} \cdot \left( k {\mitbf{\nabla} } T \right) + \tau : \epsilon , \label{entropy} \\
{\bf 0} &= - {\mitbf{\nabla} } p - \rho {\mitbf{\nabla} } \Psi + {\mitbf{\nabla} } \cdot \tau , \label{momentum} \\
0 &= \frac{\partial \rho}{\partial t} + {\mitbf{\nabla}} \cdot \left( \rho {\bf u} \right), \label{continuity} \\
{\nabla}^2 \Psi &= 4 \pi {\cal{G}} \rho, \label{gravi}
\end{align}
where $\rho$, $T$, $s$, $k$, $\tau$, $\epsilon$, $p$, $\Psi$ and ${\bf u}$ denote density, temperature, specific entropy, thermal conductivity, shear-stress tensor, rate of deformation tensor, pressure, gravitational potential and velocity fields, respectively and where ${\cal{G}}$ is the universal gravitational constant. In equation (\ref{momentum}), inertia has been neglected and the gravity field ${\bf g}$ has been written using the gravitational potential ${\bf g} = - {\mitbf{\nabla} } \Psi $.

These equations are then linearized around a state of well-mixed uniform but time dependent entropy, $\overline{s}$, hydrostatic pressure $\overline{p}$, density $\overline{\rho}$, gravity $\overline{\bf g}$ and gravitational potential $\overline{\Psi}$ depending only on radius and time, such that $\partial \overline{p}/ \partial r = -\overline{\rho}\, \overline{g}$, with $\overline{\bf g}$ satisfying the gravitational equation ${\nabla}^2 \overline{\Psi} = 4 \pi {\cal{G}} \overline{\rho}$ and $\overline{\bf g} = - {\mitbf{\nabla} } \overline{\Psi } $. Linearized variables are introduced such that $s=\overline{s}+s'$, $\rho = \overline{\rho} + \rho '$, $T = \overline{T} + \Theta$, $p = \overline{p} + p'$, $\Psi = \overline{\Psi} + \Psi '$ and ${\bf g} = \overline{{\bf g}} + {\bf g}'$. 
$\overline{T}(r)$ corresponds to an adiabatic profile, and $\Theta=T-\overline{T}(r)$ is a potential temperature.
The linearized governing equations take the form
\begin{align}
\overline{\rho}\, \overline{T} \frac{D s'}{D t} &= {\mitbf{\nabla}} \cdot \left( k {\mitbf{\nabla} } \Theta \right) + \tau : \epsilon - \overline{\rho}\, \overline{T} \frac{\partial \overline{s}}{\partial t} + {\mitbf{\nabla}} \cdot \left( k {\mitbf{\nabla} } \overline{T} \right), \label{entropy_l} \\
{\bf 0} &= - {\mitbf{\nabla} } p' - \overline{\rho} {\mitbf{\nabla} } \Psi ' - \rho ' {\mitbf{\nabla} } \overline{\Psi} + {\mitbf{\nabla} } \cdot \tau , \label{momentum_l} \\
\frac{\partial \rho '}{\partial t} &= -\frac{\partial \overline{\rho } }{\partial t}-{\mitbf{\nabla}} \cdot \left( \overline{\rho} {\bf u} \right), \label{continuity_l} \\
{\nabla}^2 \Psi ' &= 4 \pi {\cal{G}} \rho '. \label{gravi_l}
\end{align}
Using Maxwell relations, we obtain a linearized expression of $\rho '$ in terms of $s'$ and $p'$
\begin{equation}
\rho ' = \left( \frac{\partial \rho}{\partial s} \right)_{\!\!P} s' + \left( \frac{\partial \rho}{\partial P} \right)_{\!\!s} p' = - \frac{\overline{\alpha}\, \overline{\rho}\, \overline{T}}{\overline{c_p}} s'  -  \frac{1}{\overline{\rho}\, \overline{\bf g}} \frac{\partial \overline{\rho }}{\partial r} p', \label{linearrho}
\end{equation}
where $\overline{\alpha}$ and $\overline{c_p}$ are the volume expansion coefficient and specific heat capacity corresponding to the reference adiabatic state. 
With this expression for density fluctuations, equation (\ref{momentum_l}) can be written as
\begin{equation}
{\bf 0} = - \overline{\rho} {\mitbf{\nabla} } \left( \frac{p'}{\overline{\rho}} + \Psi' \right)  + \frac{\overline{\alpha}\, \overline{\rho}\, \overline{g} \, \overline{T}}{\overline{c_p}} s' \, {\bf e_r} + {\mitbf{\nabla} } \cdot \tau , \label{momentum_l2} 
\end{equation}
where ${\bf e_r}$ is the unit radial vector. {The equation of entropy fluctuations (\ref{entropy_l}) can be rewritten as
\begin{equation}
\overline{\rho}\, \frac{D\, \overline{T} s'}{D t} = - \frac{\overline{\alpha}  \overline{g} \overline{T} }{\overline{c_p}} s' u_r + {\mitbf{\nabla}} \cdot \left( k {\mitbf{\nabla} } \Theta \right) + \tau : \epsilon - \overline{\rho}\, \overline{T} \frac{\partial \overline{s}}{\partial t} + {\mitbf{\nabla}} \cdot \left( k {\mitbf{\nabla} } \overline{T} \right). \label{entropy_l2} \\
\end{equation}
}

Then, the  anelastic liquid approximation  \citep{sto01,ajs05} can be made, which consists in replacing the general  
linearized expression for entropy,
\begin{equation}
s'=  \frac{\overline{c_p}}{\overline{T}} \Theta - \frac{\overline{\alpha }}{\overline{\rho}} p', \label{linearS}
\end{equation}
by its first term only, 
\begin{equation}
s'\simeq  \frac{\overline{c_p}}{\overline{T}} \Theta, \label{ALAlinearS}
\end{equation}
under the condition $\overline{\alpha } \overline{T} Di \ll 1$ \citep{ajs05}, where $Di = \overline{\alpha} \, g_\mathrm{icb} \, r_\mathrm{ic} / \overline{c_p}$ is the dimensionless dissipation number, which compares the inner core radius $r_\mathrm{ic}$ with the natural length scale for adiabatic temperature variations, $\overline{c_p}/( \overline{\alpha} \, g_\mathrm{icb} )$. In the inner core, $Di\simeq 0.07 \times (r_\mathrm{ic}/1221\ \mathrm{km})^2$ and $\overline{\alpha}\, \overline{T} \simeq 5\ 10^{-2}$, so that the anelastic liquid approximation can be made safely. 
An alternative analysis \citep{Alboussiere2013} indicates that $c_p/c_v -1 \ll  1$, where $c_v$ is the specific heat at constant volume, is the relevant criterion for the anelastic liquid approximation. Since $c_p/c_v - 1 = \gamma \alpha T$ and the Gruneisen parameter $\gamma$ is of order
unity, this criterion is well satisfied.
Under the liquid anelastic approximation, the momentum equation (\ref{momentum_l2}) and entropy equation (\ref{entropy_l2}) can then be written as
\begin{align}
{\bf 0} &= - \overline{\rho}\, {\mitbf{\nabla} } \left( \frac{p'}{\overline{\rho}} + \Psi ' \right)  + \overline{\alpha}\, \overline{\rho}\, \overline{g}\, \Theta \, {\bf e_r} + {\mitbf{\nabla} } \cdot \tau , \label{momentum_ALA} \\
\begin{split}
\overline{\rho}\, \frac{D \left( \overline{c_p} \Theta \right) }{D t} &= - \overline{\alpha} \overline{\rho} \overline{g} \Theta u_r +    {\mitbf{\nabla}} \cdot \left( k {\mitbf{\nabla} } \Theta \right) + \tau : \epsilon \\& - \overline{\rho}\, \overline{c_p} \frac{\partial \overline{T}}{\partial t} + {\mitbf{\nabla}} \cdot \left( k {\mitbf{\nabla} } \overline{T} \right), \label{entropy_ALA}  
\end{split}
\end{align}
where terms involving $\partial \overline{c_p} / \partial t$ and $\partial \overline{\rho} / \partial t$ have been neglected in (\ref{entropy_ALA}).

The importance of self gravitation is best estimated by  analyzing its effect in terms of vorticity production.
We form the vorticity equation by taking the curl of equation \eqref{momentum_l}, which gives
\begin{align}
\mathbf{0} &=  {\mitbf{\nabla} }\bar \rho \times {\mitbf{\nabla} } \Psi' + {\mitbf{\nabla} } \rho' \times \mathbf{\bar g} + \mitbf{\nabla} \times \left( {\mitbf{\nabla} } \cdot \tau \right),\\ 
 &= \frac{d\bar\rho }{d r} {\bf e_r}  \times {\mitbf{\nabla}_{\!h} } \Psi' - \bar g\, {\mitbf{\nabla}_{\!h} } \rho' \times \mathbf{e_r} + \mitbf{\nabla} \times \left( {\mitbf{\nabla} } \cdot \tau \right),  \label{VorticityEq}
\end{align}
where $\mitbf{\nabla}_{\!h}$ denotes the horizontal part of the gradient.
The first term on the right-hand-side originates from the interaction between the mean radial density gradient and the horizontal gradient in $\Psi'$, and is to be compared with the second term, which results from the interaction between horizontal density gradients and the mean radial gravity field.
From the gravitational  equation, ${\nabla}^2 {\Psi'} = 4 \pi {\cal{G}} {\rho'}$, we find that $\Psi' \sim  4\pi {\cal{G}} {\rho'} \lambda^2$, where $\lambda$ is the typical  length scale of the temperature and gravitational potential perturbations. 
Using this estimate for $\Psi '$ the ratio of the first two terms in equation \eqref{VorticityEq} is 
\begin{equation}
\frac{\left| {\mitbf{\nabla} }\bar \rho \times {\mitbf{\nabla} } \Psi' \right|}{\left| {\mitbf{\nabla} } \rho' \times  \mathbf{\bar g} \right|} \sim \frac{d \bar \rho}{dr} \frac{\Psi'}{\bar g \rho'} \sim   \frac{d \bar \rho}{dr}\frac{4\pi \mathcal{G} \lambda^2}{\bar g}. 
\end{equation}
Noting that $d \bar \rho/dr = - (d \bar \rho/dp) \bar \rho \bar g = - \bar \rho^2 \bar g/K_S$ and that $\bar g_\mathrm{icb}=(4\pi/3)\mathcal{G}\bar \rho\, r_\mathrm{ic}$, we obtain
\begin{equation}
\frac{\left| {\mitbf{\nabla} }\bar \rho \times {\mitbf{\nabla} } \Psi' \right|}{\left| {\mitbf{\nabla} } \rho' \times \mathbf{\bar g} \right|}  \sim 3 \frac{\bar \rho \bar g_\mathrm{icb} r_\mathrm{ic}}{K_s} \frac{\lambda^2}{r_\mathrm{ic}^2} \sim 3 \frac{Di}{\gamma} \frac{\lambda^2}{r_\mathrm{ic}^2} ,
\end{equation}
where the Gr\"uneisen parameter $\gamma\simeq 1.4$ is equal to $\bar \alpha K_S/(\bar c_p \bar \rho)$.

Since $Di/\gamma \simeq 0.05$, the vorticity source arising from self-gravitation effects might be up to $\sim 15$ \% of the total vorticity production if the length scale of convection is similar to the inner core radius, but has a much smaller contribution when $\lambda/r_\mathrm{ic}$ is small. 
Although the approximation might not be very good in cases where $\lambda$ is  comparable to $r_\mathrm{ic}$, 
we will ignore here the radial variations of $\bar \rho$, without which the force arising from self-gravitation is potential, and is therefore balanced by the pressure field.
The density in the inner core is assumed to be uniform : $\overline{\rho } = \rho _s$.
To be consistent, $\overline{g}$ is assumed to be a linear function of radius, $\overline{g}=g_\mathrm{icb} r/r_\mathrm{ic}$.
Density in the liquid outer core is assumed to be uniform as well~: $\overline{\rho _l} = \rho _l$. This is not correct for the outer core as a whole, but this is an excellent approximation within the depth range of the expected topography of the inner core boundary, so that $\rho _l$ is the density of the outer core close to the inner core for our purpose. 

The rheology is assumed to be newtonian, with uniform effective viscosity $\eta$. Furthermore,  viscous and adiabatic heating can be neglected since the dissipation number is small \citep{Tritton}. 
We further assume that the thermal conductivity and thermal expansion are uniform.
With $\kappa = k/(\rho _s \, \overline{c_p})$  the thermal diffusivity, our final set of equation is
\begin{align}
{\mitbf{\nabla} } \cdot {\bf u} &= 0, \label{continuity_ALA} \\
{\bf 0} &= - {\mitbf{\nabla} } \left( {p'} + \rho_s \Psi ' \right)  +  \frac{\overline{\alpha}\, \rho_s\,g_\mathrm{icb}}{r_\mathrm{ic}}\, \Theta \,r\, {\bf e_r} + {\eta } {\mitbf{\nabla} }^2 {\bf u} , \label{momentum_ALA2} \\
 \frac{D \Theta}{D t} &= \kappa {\nabla}^2\ \Theta + S(t), \label{entropy_ALA2} 
\end{align}
where the effective heating rate $S(t)$ is defined as  the difference between secular cooling and heat conducted down the adiabat~:
\begin{equation}
S(t)= \kappa \nabla^2 \overline{T} -  \frac{\partial \overline{T}}{\partial t}.   \label{DefS}
\end{equation}
$S$ can be shown to depend mainly on time, not radius.
When this term is positive (strong secular cooling and/or weak conduction), the inner core is superadiabatic and natural convection may develop.

\subsection{Expression of boundary conditions}

Despite the fact that we have stressed the necessity for a non uniform temperature on the inner core boundary when phase changes occur (in section \ref{phase_change}), we shall now argue that the boundary condition for thermal convection within the inner core is well approximated by $\Theta = 0$ at $r=r_\mathrm{ic}$. 
Indeed, the lateral variations of potential temperature associated with the ICB dynamic topography will be found to be of order $10^{-2}$~K or smaller (corresponding to a dynamic topography $\lesssim 100$ m), while potential temperature variations within the inner core will be found to be of order $1$~K or larger. We thus assume
\begin{equation}
\Theta ( r=r_\mathrm{ic}) = 0. \label{bcT}
\end{equation}

The mechanical boundary conditions are tangential stress-free conditions (the fluid outer core cannot sustain tangential stress) and continuity of the normal stress at the inner core boundary. 
With the assumption of small topography, the normal vector is very close to the radial unit vector and the stress-free tangential conditions can be written as
\begin{align}
\tau _{r \theta} = \eta \left[ r \frac{\partial }{\partial r} \left( \frac{u_\theta }{r} \right) + \frac{1}{r } \frac{\partial u_r}{\partial \theta } \right] &= 0 , \label{stresstheta} \\
\tau _{r \phi} = \eta \left[ r \frac{\partial }{\partial r} \left( \frac{u_\phi }{r} \right) + \frac{1}{r \sin \theta } \frac{\partial u_r}{\partial \phi} \right] &= 0 , \label{stressphi} 
\end{align}
where the spherical coordinates $(r, \theta , \phi)$ are used, while the continuity of the normal stress gives
\begin{equation}
\left\llbracket 2 \eta \frac{\partial u_r}{\partial r} - p  \right\rrbracket_\mathrm{icb} = 0, \label{normalstress2}
\end{equation}
where $\llbracket \dots \rrbracket_\mathrm{icb} $ denotes the difference of a quantity across the ICB.
Using again the decomposition $p=\overline{p}+p'$, this becomes
\begin{equation}
\rho _l g_\mathrm{icb} h - p'^+ - 2 \eta \frac{\partial u_r}{\partial r} - \rho _s g_\mathrm{icb} h + p'^- = 0, \label{normalstress3}
\end{equation}
where the subscripts $+$ and $-$ denote the liquid and solid sides respectively and where overlapping adiabatic hydrostatic states have been used for the liquid and solid regions. This condition can also be written as 
\begin{equation}
- \Delta \rho\,  g_\mathrm{icb}\, h + \rho _l \Psi ' - 2 \eta \frac{\partial u_r}{\partial r} + p'^- = 0, \label{normalstress4}
\end{equation}
because integrating the hydrostatic equation in the liquid outer core leads to $p+\rho _l \Psi$ constant, which applies also to perturbation quantities. 

Finally, the radial velocity $u_r$ at the ICB is related to the topography $h$ and gravitational potential perturbation $\Psi'$ through the heat balance at the ICB [Equation \eqref{CLheat2}].

\subsection{Set of equations}

Introducing two new variables, 
\begin{align}
\hat{h} &= h + \frac{\Psi '}{g_\mathrm{icb}} , \label{hhat} \\
\hat{p} &= p'^- + \rho _s \Psi ' , \label{phat} 
\end{align}
one can write the momentum and entropy equation, together with the boundary conditions relevant when phase change is allowed between solid inner core and liquid outer core:
\begin{align} 
{\mitbf{\nabla} } \cdot {\bf u} &= 0, \label{continuity_ALAf} \\
{\bf 0} &= - {\mitbf{\nabla} } \hat{p}  + \frac{\overline{\alpha}\, \rho_s\,g_\mathrm{icb}}{r_\mathrm{ic}}\, \Theta \, r\, {\bf e_r} + {\eta } {\mitbf{\nabla} }^2 {\bf u} , \label{momentum_ALAf} \\
 \frac{D \Theta}{D t} &= \kappa {\nabla}^2  \Theta + S(t), \label{entropy_ALAf} 
\end{align}
with boundary conditions at $r=r_\mathrm{ic}$ from (\ref{bcT}), (\ref{stresstheta}), (\ref{stressphi}), (\ref{normalstress4}) and (\ref{CLheat2}) : 
\begin{align} 
  \Theta&=0 , \label{bcTf} \\
  \tau _{r \theta} = \eta \left[ r \frac{\partial }{\partial r} \left( \frac{u_\theta }{r} \right) + \frac{1}{r } \frac{\partial u_r}{\partial \theta } \right] &= 0 , \label{stressthetaf} \\
  \tau _{r \phi} = \eta \left[ r \frac{\partial }{\partial r} \left( \frac{u_\phi }{r} \right) + \frac{1}{r \sin \theta } \frac{\partial u_r}{\partial \phi} \right] &= 0 , \label{stressphif} \\
  - \Delta \rho\,  g_\mathrm{icb}\, \hat{h}  - 2 \eta \frac{\partial u_r}{\partial r} + \hat{p} &= 0, \label{normalstressf}\\
  u_r - \dot r_\mathrm{ic}  &= \frac{ \hat{h} }{\tau _\phi }. \label{CLheatf} 
\end{align}
It can be seen from (\ref{normalstressf}) and (\ref{CLheatf}), that $\hat{h}$ is not necessary for the resolution of the equations, although it can be recovered once the problem is solved,  and can be eliminated between these two equations, leaving only one boundary condition:
\begin{equation}
- \Delta \rho  g_\mathrm{icb} \tau _\phi  ( u_r - \dot r_\mathrm{ic})   - 2 \eta \frac{\partial u_r}{\partial r} + \hat{p} = 0, \label{normalstressf2}
\end{equation}
Incidently, it can also be seen that there is no  need to explicitly solve the gravitational equation (\ref{gravi_l}), since $\Psi '$ has been absorbed in the modified pressure (\ref{phat}). 

The governing equations and boundary conditions are now made dimensionless using the age of the inner core $\tau _{ic}$, its time dependent radius $r_\mathrm{ic} (t)$, $\kappa / r_\mathrm{ic} (t)$, $\eta \kappa / r_\mathrm{ic}^2 (t)$ and $S(t) r_\mathrm{ic}^2 (t) / (6 \kappa )$ as scales for time, length, velocity, pressure and potential temperature respectively. Using the same symbols for dimensionless quantities, dimensionless equations can be written as
\begin{align} 
{\mitbf{\nabla} } \cdot {\bf u} &= 0, \label{continuity_ALAd} \\
{\bf 0} &= - {\mitbf{\nabla} } \hat{p}  + Ra(t)\, \Theta\, {\bf r} + {\mitbf{\nabla} }^2 {\bf u} , \label{momentum_ALAd} \\
\begin{split}
\xi (t) \frac{\partial \Theta}{\partial  t} &= {\nabla}^2  \Theta - ({\bf u} -Pe (t) {\bf r} ) \cdot {\mitbf{\nabla}} \Theta\\ 
&+ 6 - \left( \xi (t) \frac{\dot{S} (t) \tau _{ic}}{S (t) } + 2 \, Pe (t) \right) \Theta, \label{entropy_ALAd} 
\end{split}
\end{align} 
where ${\bf r} = r\, \bf e_r$ and  $\dot{S} (t) = d S(t) / dt $.
The last terms in (\ref{entropy_ALAd}) are due to dependency of the temperature scale on time, when used to make the equations dimensionless.  Three dimensionless parameters are needed
\begin{align}
\xi (t) &= \frac{r_\mathrm{ic}^2 (t)}{\kappa \tau _{ic}}, \label{zeta} \\
Pe (t) &= \frac{r_\mathrm{ic} (t) \dot r_\mathrm{ic} (t) }{\kappa}, \label{peclet} \\
Ra (t) &= \frac{\alpha \rho _s g_\mathrm{icb}(t) S(t) r_\mathrm{ic}^5 (t) }{6 \kappa ^2 \eta }. \label{rayleigh}
\end{align}
The dimensionless boundary conditions, at $r = 1$, can be written
\begin{align} 
 \Theta&=0 , \label{bcTfa} \\
 \tau _{r \theta} =  r \frac{\partial }{\partial r} \left( \frac{u_\theta }{r} \right) + \frac{1}{r } \frac{\partial u_r}{\partial \theta }  &= 0 , \label{stressthetafa} \\
 \tau _{r \phi} =  r \frac{\partial }{\partial r} \left( \frac{u_\phi }{r} \right) + \frac{1}{r \sin \theta } \frac{\partial u_r}{\partial \phi}  &= 0 , \label{stressphifa} \\
 - {\cal{P}} (t) ( u_r - \dot r_\mathrm{ic})  - 2  \frac{\partial u_r}{\partial r} + \hat{p} &= 0, \label{normalstressfa}   
\end{align}
where we have introduced the "phase change number", $\mathcal{P}$, characterizing the resistance to phase change :
\begin{equation}
{\cal{P}} (t) = \frac{\Delta \rho \, g_\mathrm{icb} (t) \, r_\mathrm{ic} (t) \, \tau _\phi(t) }{\eta }. \label{P}
\end{equation}
$\mathcal{P}$ is the ratio between the phase change timescale $\tau_\phi$ and the viscous relaxation timescale $\tau_\eta = \eta/(\Delta \rho \, g_\mathrm{icb} \, r_\mathrm{ic})$ (equivalent to post-glacial rebound timescale).
$\mathcal{P}=0$ corresponds to instantaneous melting or freezing, while $\mathcal{P}\rightarrow \infty$ corresponds to infinitely slow melting or freezing.
In the limit of infinite $\mathcal{P}$, the boundary condition \eqref{normalstressfa} reduces to the condition $u_r=0$, which corresponds to impermeable conditions. 
In contrast, when $\mathcal{P}\rightarrow 0$, Equation \eqref{normalstressfa} implies that the normal stress tends toward 0 at the boundary, which corresponds to fully permeable boundary conditions. 
The general case of finite $\mathcal{P}$ gives boundary conditions for which the rate of phase change at the boundary (equal to $u_r$) is proportional to the normal stress induced by convection within the spherical shell.

A steady state version of the set of equation \eqref{continuity_ALAd}-\eqref{P} is found by using $r_\mathrm{ic}^2/\kappa$ as a timescale instead of $\tau_\mathrm{ic}$, and keeping  $r_\mathrm{ic}$ and $S$ constant.
All the equation remain unchanged except the heat equation which now writes
\begin{equation} 
\frac{\partial \Theta}{\partial  t} = {\nabla}^2  \Theta - {\bf u}  \cdot {\mitbf{\nabla}} \Theta + 6.  \label{EqPotentialTemperatureSteadyState}
\end{equation} 
This will be used in section \ref{Numerical_results} where numerical simulations with constant inner core radius and thermal forcing will be used to derive scaling laws.

With the assumptions made so far, the velocity field is known to be purely poloidal \citep{Ribe07}, and we introduce the poloidal scalar $P$ defined such that
\begin{equation}
{\bf u} = {\mitbf{\nabla} } \times {\mitbf{\nabla} } \times \left( P {\bf r} \right). \label{poloidal}
\end{equation}
Taking the curl of the momentum equation \eqref{momentum_ALAd} gives
\begin{equation}
Ra (t) L^2 \Theta = \left( {\nabla }^2 \right) ^2 L^2 P , \label{poloidalP}
\end{equation}
where the angular momentum operator $L^2$ is
\begin{equation}
L^2 = -  \frac{1}{\sin \theta} \frac{\partial}{\partial \theta} \left( \sin \theta  \frac{\partial}{\partial \theta} \right) - \frac{1}{\sin ^{2} \theta}\frac{ \partial^2}{\partial \phi^2}.
\end{equation}
Horizontal integration of the momentum equation (see \cite{FP87,Ribe07}, where this is done component-wise in spherical harmonics) shows that, on $r=1$
\begin{equation}
- \hat{p} + \frac{\partial }{\partial r } \left( r {\nabla}^2 P  \right) = C^{st}. \label{horizontal}
\end{equation}
This expression can be used to eliminate $\hat{p}$ in the boundary condition \eqref{normalstressfa}.
Noting that
\begin{equation}
u_r = \frac{1}{r}  L^2 P,
\end{equation}
continuity of the normal stress at the ICB  (equation \eqref{normalstressfa}) gives the following boundary condition at $r=1$:
\begin{equation}
\frac{\partial }{\partial r} \left( r {\nabla}^2 P  - \frac{2}{r} L^2 P \right) -  {\cal{P}} (t)\, \left( \frac{1}{r} L^2 P - \dot r_\mathrm{ic}\right)  = C^{st}, \label{normalstressf2a}
\end{equation}
while the stress-free conditions (\ref{stressthetafa}) and (\ref{stressphifa}) take the form
\begin{equation}
r \frac{\partial }{\partial r} \left( \frac{1}{r^2} \frac{\partial }{\partial r} \left( r P \right)   \right) + \frac{1}{r^2}L^2 P = C^{st},
\end{equation}
which can be rewritten as
\begin{equation}
\frac{\partial^2 P}{\partial r^2} + \left( L^2 -{2} \right) \frac{P}{r^2} = C^{st}. \label{stressfreeP}
\end{equation}

At this stage, there are two unknown scalar field variables, $\Theta$ and $P$. They are expanded as
\begin{eqnarray}
\Theta &=& t_l^m (r,t)\, Y_l^m, \label{harmonyT} \\
P &=& p_l^m (r,t)\, Y_l^m, \ \ \ \ l \geq 1, \label{harmonyP}
\end{eqnarray}
where $Y_l^m (\theta, \phi ) $, for $l \geq 0$, $m \in [-l ; l]$  are surface spherical harmonics, which satisfy $L^2 Y_l^m = l(l+1) Y_l^m$. 
The momentum equation (\ref{poloidalP}) takes the form 
\begin{equation}
Ra (t) t_l^m = \mathcal{D}_l^2 p_l^m , \label{poloidalPlm}
\end{equation}
where
\begin{equation}
\mathcal{D}_l = \frac{d^2 }{d r^2} + \frac{2}{r} \frac{d}{d r} - \frac{l(l+1)}{r^2}. \label{Dl}
\end{equation}
The stress-free boundary condition (\ref{stressfreeP}) can be written as
\begin{equation}
\frac{d^2 p_l^m }{d r^2} + \left[ l(l+1) -2 \right] \frac{p_l^m}{r^2} = 0, \ \ \ \ l \geq 1, \label{stressfreePlm}   
\end{equation}
and the boundary condition (\ref{normalstressf2a}), derived from normal stress balance, as
\begin{equation}
\frac{d }{d r} \left(r \mathcal{D}_l p_l^m -2l(l+1) \frac{p_l^m}{r}   \right) = l(l+1) {\cal{P}} (t) \frac{p_l^m}{r} , \ \ \ \ l \geq 1. \label{normalstressPlm}
\end{equation}
With (\ref{stressfreePlm}), the equation above can also be written: 
\begin{equation}
r^2 \frac{d^3 p_l^m }{d r^3} -3 l (l+1)  \frac{d p_l^m }{d r} = \left[ l(l+1) {\cal{P}} (t) -\frac{6}{r} \right] {p_l^m} , \ \ \ \ l \geq 1. \label{normalstressPlm2}
\end{equation}
The thermal equation is also written in spherical harmonic expansion but cannot be solved independently for each degree and order due to the non-linearity of the advection term, which is evaluated in the physical space and expanded back in spherical harmonics. 

\section{Linear stability analysis}
\label{linear_stability}

We investigate here the linear stability of the system of equations  describing thermal convection in the inner core with phase change at the ICB, as derived in section \ref{gov_eqns}. 
The calculation given here is a generalization of the linear stability analysis of thermal convection in an internally heated sphere given by \cite{Chandrasekhar1961}.
The case considered by \cite{Chandrasekhar1961}, where a non-deformable, impermeable outer boundary is assumed, corresponds to the limit $\mathcal{P}\rightarrow\infty$ of the problem considered here.

We assume constant $Ra$ and $\mathcal{P}$ (and $\xi=1$, $Pe=\dot S = 0$),
 thus ignoring that the base diffusive solution itself is time-dependent. 
 This assumption is essentially correct when the growth rate of the fastest unstable disturbance is much larger than the growth rate of the radius of the inner core. 
The basic state of the problem is then given by
\begin{align}
\bar \Theta &= 1 - r^2, \label{Tbase} \\  
\bar{ \mathbf{u}} &= \mathbf{0},
\end{align}
which is the steady conductive solution of the system of equation developed in section \ref{gov_eqns}.
We investigate the stability of this conductive state against infinitesimal perturbations of the temperature and velocity fields.
The temperature field is written as the sum of the conductive temperature profile given by equation \eqref{Tbase} and infinitesimal disturbances $\tilde \Theta$, $\Theta(r,\theta,\phi,t) = \bar \Theta(r) + \tilde \Theta(r,\theta,\phi,t)$. 
The velocity field perturbation is noted $\tilde{\mathbf{u}}(r,\theta,\phi,t)$, and has an associated poloidal scalar $\tilde P(r,\theta,\phi,t)$.
We expand the temperature and poloidal disturbances in spherical harmonics,
\begin{align}
\tilde \Theta &= \sum_{l=0}^\infty \sum_{m=-l}^l \tilde t_l^m(r) Y_l^m(\theta,\phi)\, \mathrm{e}^{\sigma_l t},   \label{TDecomposition}\\
\tilde P &=\sum_{l=1}^\infty \sum_{m=-l}^l \tilde p_l^m(r) Y_l^m(\theta,\phi)\, \mathrm{e}^{\sigma_l t}, \label{PDecomposition}
\end{align}
where $\sigma_l$ is the growth rate of the degree $l$ perturbations (note that since $m$ does not appear in the  system of equations, the growth rate is function of $l$ only, not $m$). 

The only non-linear term in the system of equations is the advection of heat $\mathbf{u}\cdot \nabla \Theta$ in Equation \eqref{EqPotentialTemperatureSteadyState}, which is linearized as 
\begin{equation}
\tilde u_r \frac{\partial \bar \Theta}{\partial r} = -2 r \tilde u_r = - 2 L^2 \tilde P .  
\end{equation}
The resulting linearized transport equation for the potential temperature disturbance is
\begin{equation} 
\left(\frac{\partial }{\partial  t} -  {\nabla}^2 \right)\tilde \Theta=  2 L^2 \tilde P + 6.
\end{equation} 
Using the decompositions \eqref{TDecomposition} and \eqref{PDecomposition}, the linearized system of equations is then, for $l\geq 1$,
\begin{align}
Ra\, \tilde t_l^m &= \mathcal{D}_l^2 \tilde p_l^m,\\
\left( \sigma_l -  \mathcal{D}_l \right) \tilde t_l^m &= 2 l (l+1) \tilde p_l^m ,  \label{HeatMarg}
\end{align}
with the boundary conditions given by Equations \eqref{stressfreePlm} and \eqref{normalstressPlm}, with $\tilde t_l^m(r=1)=0$.

Developing the $\tilde t_l^m$  in series of spherical Bessel functions and solving for $\tilde p_l^m$, we obtain an infinite set of linear equations in perturbation quantities, which admits a non trivial solution only if its determinant is equal to zero (see Appendix \ref{AppendixLinearStability} for the details of the calculation). This provides the following dispersion equation, 
\begin{equation}
\begin{split}
&\left|\left| \left[  q_3^l(\mathcal{P}) \alpha_{l,i}^2 +   q_4^l(\mathcal{P}) \right] \left[1-\frac{4l+6}{\alpha_{l,j}^2} \right] -\left[ q_1^l(\mathcal{P}) \alpha_{l,i}^2 +  q_2^l(\mathcal{P}) \right]  \right.\right. \\
+& \left.\left. \left( \frac{\sigma_l \alpha_{l,i}^4 + \alpha_{l,i}^6}{2l(l+1) Ra}  - 1\right) \frac{1}{2}   \delta_{ij} \right|\right| =0,  \quad \text{with}\ \ i,j=1,2,...
\end{split}\label{Determinant_l_text}
\end{equation}
where $||...||$ denotes the determinant.
Here $\alpha_{l,i}$ denotes the $i$th zero of the spherical Bessel function of degree $l$. The functions $q_1^l(\mathcal{P})$ to $q_4^l(\mathcal{P})$ are given in Appendix \ref{AppendixLinearStability} by equations \eqref{q1}, \eqref{q2}, \eqref{q3} and \eqref{q4}.

\begin{figure}
\begin{center}
\includegraphics[width=\columnwidth]{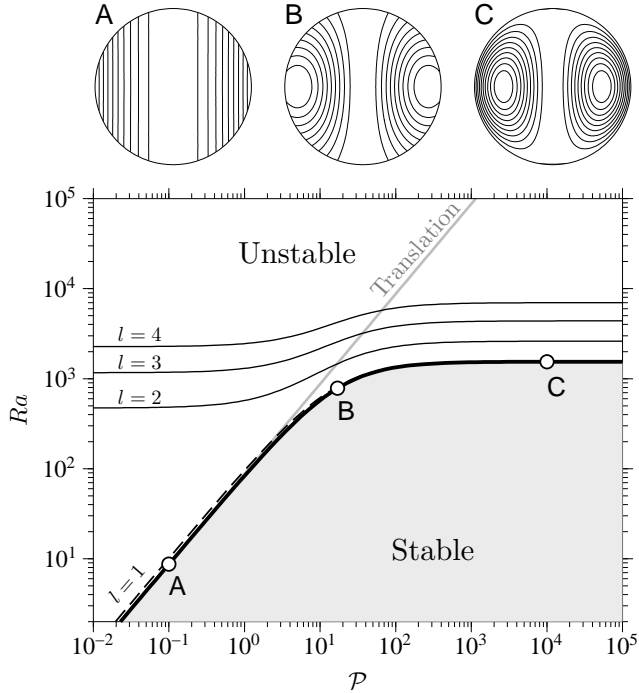}
\end{center}
\caption{Stability diagram for convection in a sphere with phase change at its outer boundary.
The neutral stability curve ($l=1$) obtained by solving equation \eqref{Determinant_1} with $\sigma_1=0$ is shown by the thick black line. The dashed line shows the approximate stability curve given by equation \eqref{RelationDispersionApprox}.
The neutral stability curves of higher modes ($l=2,3,4$) obtained by solving equation \eqref{Determinant_l_text} with $\sigma_l=0$ are shown by the annotated thin black lines.
The neutral stability curves for $l\geq5$ are not shown to avoid overcrowding the figure.
The thick grey curve annotated "Translation" is the neutral stability curve of the translation mode, given by equation \eqref{Critical_RaP}.
Streamlines of the first unstable mode at points $\mathsf{A}$ ($\mathcal{P} = 0.1$), $\mathsf{B}$ ($\mathcal{P} = 17$), and $\mathsf{C}$ ($\mathcal{P} = 10^4$) are shown in the upper figure. 
\label{FigStabDiag}
}
\end{figure}

Solving equation \eqref{Determinant_l_text} for a given value of $l$ and $\sigma_l=0$ gives the critical value $Ra_c$ of the Rayleigh number for instability of the degree $l$ mode, as a function of $\mathcal{P}$.
The resulting marginal stability curves for $l=1$ to 4 are shown in figure \ref{FigStabDiag}.
The first unstable mode is always the $l=1$ mode, for which equation \eqref{Determinant_l_text} reduces to 
\begin{equation}
\left|\left| \left(Ra - \frac{\alpha_{1,i}^6+\sigma_1 \alpha_{1,i}^4}{4 } \right)  \delta_{ij}  +   {\alpha_{1,i}^2}\frac{Ra}{\mathcal{P} }  +\frac{20}{3\, \alpha_{1,j}^2 }Ra \right|\right| = 0. \label{Determinant_1}
\end{equation}
A useful first approximation is obtained by keeping only the $i=j=1$ terms, thus setting the $(1,1)$ component of the matrix to zero.
This gives a simple analytical form for the growth rate,
\begin{equation}
\sigma_1 =  \left( \frac{4}{\alpha_{1,1}^4} + \frac{80}{3\, \alpha_{1,1}^6}  \right) Ra + \frac{4}{\alpha_{1,1}^2} \frac{Ra}{\mathcal{P}} - \alpha_{1,1}^2
\end{equation}
and for the critical Rayleigh number,
\begin{equation}
   Ra_c = \frac{\alpha_{1,1}^6}{4 } \left[  1  + \frac{\alpha_{1,1}^2}{\mathcal{P} }  + \frac{20}{3} \frac{1}{\alpha_{1,1}^2}  \right]^{-1}, \label{RelationDispersionApprox}
\end{equation}
with $\alpha_{1,1}\simeq 4.4934$.
When $\mathcal{P}\ll 1$ or $\mathcal{P}\gg 1$, $Ra_c$ and $\sigma_l$ have the following limits :
\begin{align}
Ra_c &\rightarrow 
\begin{cases}
\dfrac{\alpha_{1,1}^8}{4\alpha_{1,1}^2 + 80/3 } \simeq 1547\hspace*{0.9cm}  &\text{when}\ \mathcal{P}\rightarrow \infty, \\
\dfrac{\alpha_{1,1}^4}{4 } \mathcal{P} \simeq 101.9\,  \mathcal{P} &\text{when} \ \mathcal{P}\rightarrow 0,  \label{Critical_RaP_approx}
\end{cases}  \\
\sigma_1 &\rightarrow 
\begin{cases}
 \left(\dfrac{4}{\alpha_{1,1}^4} + \dfrac{80}{3\, \alpha_{1,1}^6}  \right) Ra  - \alpha_{1,1}^2\  &\text{when}\ \mathcal{P}\rightarrow \infty, \\
\dfrac{4}{\alpha_{1,1}^2} \dfrac{Ra}{\mathcal{P}} - \alpha_{1,1}^2\ &\text{when} \ \mathcal{P}\rightarrow 0,   \label{GrowthRates}
\end{cases}
\end{align}
Higher order approximations can be obtained by retaining more terms in the determinant. 
For $\mathcal{P}\gg 1$, the critical value of $Ra$ converges toward
\begin{equation}
Ra_c=1545.6,  \label{Rac_largeP}
\end{equation}
in agreement with \cite{Chandrasekhar1961}'s result (the value given by \cite{Chandrasekhar1961} is twice the value given here, because of different definitions of $Ra$).
When  $\mathcal{P}\ll 1$, the relevant non-dimensional parameter is the ratio $Ra/\mathcal{P}$, which is independent of the viscosity and of the thermal diffusivity.
An exact value of the critical value of $Ra/\mathcal{P}$ will be given below (equation \eqref{Critical_RaP}).

The pattern of the first unstable mode can be calculated by solving the system \eqref{SystemEquationsLS} given in Appendix \ref{AppendixLinearStability} for given $\mathcal{P}$ and $Ra$.	
The first unstable modes calculated in this way for points $\mathsf{A}$, $\mathsf B$ and $\mathsf C$ ($\mathcal{P}=0.1$, 17 and $10^4$) in the stability diagram are shown in  figure \ref{FigStabDiag}. 
As shown in Appendix \ref{AppendixLinearStability}, the $l=1$, $m\in [-1,0,1]$ components of the poloidal scalar can be written as
 \begin{equation}
 \tilde{p}_1^m = \sum_{i=1}^\infty A_{1,i} \left[  \frac{ j_1(\alpha_{1,i} r)}{\alpha_{1,i}}  +  \frac{j_2( \alpha_{1,i})}{3}( r - r^{3}) + \frac{ j_2( \alpha_{1,i})\alpha_{1,i}^2}{2\, \mathcal{P}} r \right],  \label{p_10}
 \end{equation}
where the coefficients $A_{1,i}$ are found by solving the system of equations \eqref{SystemEquationsLS}. Here $j_1$ and $j_2$ denote the spherical Bessel functions of the first kind of order 1 and 2, respectively.
From equation \eqref{p_10}, it can be seen that 
\begin{equation}
\tilde{p}_1^m \rightarrow \left( \sum_{i=1}^{\infty} A_{1,i}\,  j_2( \alpha_{1,i})\alpha_{1,i}^2   \right) \frac{2}{\mathcal{P}}r\quad \text{when} \ \mathcal{P}\rightarrow 0,
\end{equation}
which corresponds to a translation (it can be verified that a $l=1$ flow with $p_1^m \propto r$ corresponds to a flow with uniform velocity).
This is the dominant mode when $\mathcal{P}$ is small, as illustrated in figure \ref{FigStabDiag} (point $\mathsf{A}$, $\mathcal{P}=0.1$).
There is no deformation associated with this mode.

At high $\mathcal{P}$, the term in $1/\mathcal{P}$ in equation \eqref{p_10} becomes negligible, and the  first unstable mode is identical to the classical single cell degree one  mode of thermal convection with shear-free boundary and no phase change \citep{Chandrasekhar1961}, as illustrated in figure \ref{FigStabDiag} (point $\mathsf{C}$, $\mathcal{P}=10^4$).
There is no melting or solidification associated with this mode, which is apparent from the fact that the streamlines of the flow are closed.
At intermediate values of $\mathcal{P}$, the first unstable mode is a linear combination of the high-$\mathcal{P}$ convection mode and of the small-$\mathcal{P}$ translation mode.

Allowing only for the translation  (\textit{i.e.} keeping only the $p_{1,i} \propto r/\mathcal{P}$ terms), the dispersion relation \eqref{Determinant_1} reduces to
\begin{equation}
\left|\left|  \delta_{ij}  - \frac{4}{\alpha_{1,i}^4} \frac{Ra}{\mathcal{P} } \right|\right| = 0.
\end{equation}
Using Sylvester's  determinant theorem, we find that  
\begin{equation}
\left|\left|  \delta_{ij}  - \frac{4}{\alpha_{1,i}^4} \frac{Ra}{\mathcal{P} } \right|\right| = 1 - 4 \frac{Ra}{\mathcal{P} }  \sum_{i=1}^\infty \frac{1}{\alpha_{1,i}^{4}},
\end{equation}
which allows to write the critical value of $Ra/\mathcal{P}$ as
\begin{equation}
\left(\frac{Ra}{\mathcal{P}}\right)_{\!\!c} =\frac{1}{4} \left( \sum_{i=1}^\infty \frac{1}{\alpha_{1,i}^{4}} \right)^{-1} = \frac{175}{2}=87.5,  \label{Critical_RaP}
\end{equation}
where we have used \cite{sneddon1960}'s result that $\sum_{i=1}^\infty \alpha_{1,i}^{-4}=1/350$.
The critical value 175/2 is exact, and is to be preferred to the approximate value (101.9) obtained in equation \eqref{Critical_RaP_approx}.
Equation \eqref{Critical_RaP} gives the marginal stability curve shown in grey in figure \ref{FigStabDiag}.
Although the translation mode can be unstable at all value of $\mathcal{P}$ provided that $Ra$ is large enough, it is apparent in figure \ref{FigStabDiag} that the  one cell convection mode is the first unstable mode whenever $\mathcal{P}$ is larger than $\simeq Ra_c/(Ra/\mathcal{P})_c \simeq 17$ (point $\mathsf{B}$ in figure \ref{FigStabDiag}).

Finally, it can be seen in figure \ref{FigStabDiag} that the critical Rayleigh number $Ra_c^l$ for higher order modes ($l>1$) is also lowered when $\mathcal{P}\lesssim17$. However, the decrease in $Ra_c^l$ is not as drastic as it is for the $l=1$ mode because, whatever the value of $\mathcal{P}$, viscous dissipation always limits the growth of these modes.
The effect of $\mathcal{P}$ on  $Ra_c^l$ becomes increasingly small as $l$ increases.
This suggests that allowing for phase change at the ICB would generally enhance large scale motions at the expense of smaller scale motions.

\section{Analytical solutions for small $\mathcal{P}$}
\label{Small_P}

We now search for a finite amplitude solution of inner core convection at small $\mathcal{P}$.
In the limit of infinite viscosity ($\mathcal{P}\rightarrow0$), the only possible motions of the inner core are rotation, which we don't consider here, and translation. 
Guided by the results of the linear stability analysis, we search for a solution in the form of a translation.
\cite{Alboussiere2010} found a solution for the velocity of inner core translation from a global force balance on the inner core, under the assumption that the inner core is rigid. 
One of the goal of this section is to verify that the system of equations developed in section \ref{gov_eqns} indeed leads to the same solution when $\mathcal{P}\rightarrow 0$.

If the viscosity is taken as infinite and $\mathcal{P}$ is formally put to zero, 
searching for a pure translation solution and ignoring any deformation in the inner core leads to an undetermined system.
Translation is an exact solution of the momentum equation, but the translation rate is left undetermined, because all the terms in the boundary conditions \eqref{stressthetafa}, \eqref{stressphifa}, \eqref{normalstressfa} (zero tangential stress and continuity of the normal stress) vanish.
This of course does not mean that the stress magnitude vanishes, but rather that the rheological relationship between stress and strain via the viscosity becomes meaningless if the viscosity is assumed to be infinite. 
The ICB topography associated with the translation is sustained by the non-hydrostatic stress field which, even if $\eta \rightarrow \infty$, must  remain finite.
One way to calculate the stress field is to evaluate the flow induced  by the lateral temperature variations associated with the translation, for small but non-zero $\mathcal{P}$, and then take the limit $\mathcal{P}\rightarrow 0$.
If only the 'rigid inner core' limit is wanted, it suffices to calculate the flow at $\mathcal{O}(\mathcal{P})$.
The effect of finite viscosity on the translation mode can be estimated by calculating the velocity field at a higher order in $\mathcal{P}$.

\subsection{Translation velocity at zeroth order in $\mathcal{P}$}

Noting $V_0$ the translation velocity at zeroth order in $\mathcal{P}$, the poloidal scalar takes the form
\begin{equation}
p_1^0 = \frac{V_0}{2} \ r, \label{p10}
\end{equation}
with $Y_1^0 =  \cos \theta$, in a cylindrical coordinate system of axis parallel to the velocity translation. 
If $V_0$ is large enough, the temperature equation (\ref{entropy_ALAd}) has a fast convective solution whereby ${\bf u} \cdot {\mitbf{\nabla}} \Theta$ balances the constant $6$. 
Imposing $\Theta=0$ at the ICB on the crystallizing hemisphere, and ignoring a thin boundary layer below the ICB on the melting hemisphere, the temperature field, shown in figure \ref{Analytical_T_Vorticity},  is
\begin{equation}
\Theta = \frac{6}{V_0} \left( r\, \cos \theta + \sqrt{1 - r^2 \sin^2 \theta}  \right)   \label{TemperatureTranslation}
\end{equation}
\citep{Alboussiere2010}.
This results in a uniform temperature gradient in the translation direction, with the $l=1$, $m=0$ component of the temperature field being
\begin{equation}
t_1^0 = \frac{6}{V_0} \ r. \label{t10}
\end{equation}
This temperature field induces a secondary $l=1,m=0$ flow which must vanish when $\mathcal{P}\rightarrow0$. We therefore write $p_1^0$ as
\begin{equation}
p_1^0 = \frac{V_0}{2}\left[ r + \hat{p}_{1,1}^0 \mathcal{P} + \mathcal{O}(\mathcal{P}^2) \right].
\end{equation}
Inserting this form for $p_1^0$ and the temperature degree one component $t_1^0$ into the momentum equation (\ref{poloidalPlm}) gives
\begin{equation}
12 \frac{Ra}{\mathcal{P}} \frac{1}{V_0^2} r = \mathcal{D}_1^2 \hat{p}_{1,1}^0,   \label{Momentum_f}
\end{equation}
from which we can already infer that
\begin{equation}
V_0 \sim \sqrt{\frac{Ra}{\mathcal{P}}}.
\end{equation}
Equation \eqref{Momentum_f} has a general solution of the form
\begin{equation}
\hat{p}_{1,1}^0 = A r + B r^3 + C r^5, \label{p10gene}
\end{equation}
where $A$, $B$ and $C$ are constants to be determined. 

From the momentum equation (\ref{Momentum_f}), we obtain
\begin{equation}
C = \frac{3}{70} \frac{Ra}{\mathcal{P}}\frac{1}{V_0^2} . \label{CoeffC}
\end{equation}
The stress-free boundary condition (\ref{stressfreePlm}) for a degree one component,
\begin{equation}
\left.\frac{d^2 p_1^0 }{d r^2}\right|_{r=1} = 0,  \label{stressfreeP10}  
\end{equation}
leads to
\begin{equation}
B = - \frac{10}{3} C = -   \frac{1}{7} \frac{Ra}{\mathcal{P}} \frac{1}{V_0^2}  . \label{CoeffB}
\end{equation}
Finally, the condition of continuity of the normal stress (\ref{normalstressPlm2}) leads to
\begin{equation}
\left( -3 B + 18 C - 1 \right) \mathcal{P} - 2 \mathcal{P}^2 (A+B+C) + \mathcal{O}(\mathcal{P}^2 ) = 0,
\end{equation}
which implies that
\begin{equation}
-3 B + 18 C - 1 = 0. \label{NormalStress_BC}
\end{equation}
Note that the constant $A$ is left undetermined : considerations of the velocity field at order $\mathcal{P}^2$ and of the temperature field at order $\mathcal{P}$ are required to determine it.
With $B$ and $C$ given by equations \eqref{CoeffB} and \eqref{CoeffC}, equation \eqref{NormalStress_BC} gives the translation velocity $V_0$ as
\begin{equation}
V_0  = \sqrt{\frac{6}{5} \frac{Ra }{{\cal{P}}}}. \label{V2approx}
\end{equation}
In dimensional unit, the translation rate is given by
\begin{equation}
\frac{\kappa}{r_\mathrm{ic}}\sqrt{\frac{6}{5} \frac{Ra }{{\cal{P}}}} =  \left( \frac{1}{5} \frac{\rho_s}{\Delta \rho} \frac{\alpha   S }{ \tau _\phi} \right)^{1/2} r_\mathrm{ic}
\end{equation}
which, with $\tau_\phi$ given by equation \eqref{tauphi},  is exactly the same solution as that found in \cite{Alboussiere2010} from an analysis of the global force balance on the inner core.
As expected, the translation rate is independent of the inner core viscosity $\eta$ and of the thermal diffusivity, and is an increasing function of the heating rate $S$ and a decreasing function of the phase change time scale. 

The potential temperature difference across the inner core is $12/V_0$ in non-dimensional units, and
\begin{equation}
12 \frac{S r_\mathrm{ic}^2}{6 \kappa} \left( \frac{5}{6} \frac{\mathcal{P}}{Ra} \right)^{1/2} =  \left( 20 \frac{\Delta \rho\, \tau _\phi\, S}{\rho_s\, \alpha} \right)^{1/2} 
\end{equation}
in dimensional units.

\subsection{Translation velocity at $\mathcal{O}(\mathcal{P})$}
\label{SectionTranslationOrdreP}

The translation velocity at $\mathcal{O}(\mathcal{P})$ can be obtained by calculating the temperature field at $\mathcal{O}(\mathcal{P})$ and the velocity field at $\mathcal{O}(\mathcal{P}^2)$, which allows to determine the constant $A$ in the expression of the poloidal scalar at $\mathcal{O}(\mathcal{P})$ (equation \eqref{p10gene}).
The procedure, detailed in Appendix \ref{Appendix_Tr}, is  complicated by the non-linearity of the heat equation~: coupling of higher order harmonics component of the temperature and velocity fields contribute to the $l=1$ component of the temperature field.
Taking into account the effect of the non-linear coupling of the $l=2$ components of the temperature and velocity fields, we obtain
\begin{equation}
V = \sqrt{\frac{6}{5} \frac{Ra }{{\cal{P}}}} \left[ 1 - 0.0216\, \mathcal{P} + \mathcal{O}(\mathcal{P}^2)   \right]. \label{TranslationOrdreP}
\end{equation}
which suggests that the effect of deformation becomes important when $\mathcal{P}$ is a significant fraction of $1/0.0216 \simeq 46$, in agreement with the prediction of the linear stability analysis.

The temperature field and the $\phi$-component of the vorticity field at $\mathcal{O}(\mathcal{P})$, as calculated in Appendix \ref{Appendix_Tr}, are shown in figure \eqref{Analytical_T_Vorticity}.
\begin{figure}
\begin{center}
\includegraphics[width=\columnwidth]{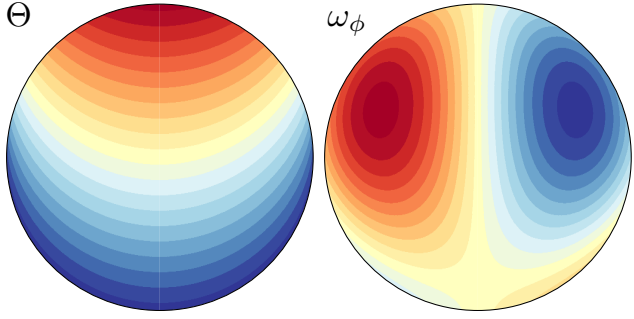}
\end{center}
\caption{Temperature field (left, red=hot, blue=cold) and vorticity field at $\mathcal{O}(\cal{P})$ (right, blue=negative, red=positive) in a meridional  cross-section (the direction of translation is arbitrary).
\label{Analytical_T_Vorticity}
}
\end{figure}

\subsection{The effect of the boundary layer}
\label{SectionBoundaryLayer}

Let us finally discuss the influence of the  thermal boundary layer that must develop in the solid inner core near the melting side when a convective translation exists. From the thermal equation (\ref{entropy_ALAd}), and with the boundary condition (\ref{bcTfa}), a thermal boundary layer of thickness $V^{-1}$ results from the balance between convective and diffusive terms, so that the degree one temperature component (\ref{t10}) may be approximated by
\begin{equation}
t_1^0  \simeq  \frac{6}{V}  \left( r - e^{   V (r-1)  } \right)  . \label{t10mod}
\end{equation}
We now note that
\begin{equation}
\mathcal{D}_1 \left( \mathrm{e}^{V(r-1)} \right) = \left(  1 + \frac{2}{V r} - \frac{2}{V^2 r^2}  \right) V^2 \mathrm{e}^{V(r-1)},
\end{equation}
so that to a good approximation,
\begin{equation}
\mathcal{D}_1 \left( \mathrm{e}^{V(r-1)} \right) \simeq V^2 \mathrm{e}^{V(r-1)}\quad \text{and}\quad \mathcal{D}_1^2 \left( \mathrm{e}^{Vr} \right) \simeq V^4 \mathrm{e}^{V(r-1)}
\end{equation}
when $V\gg1$.
Under this assumption, the resulting general solution for the velocity poloidal component (\ref{p10gene}) becomes
\begin{equation}
p_1^0 \simeq \frac{V_0}{2}\left[ r + \left(Ar + B r^3 + C r^5 - 10 \frac{V_0^2}{V^6} \mathrm{e}^{V (r-1)} \right) \mathcal{P}+ \mathcal{O}(\mathcal{P}^2) \right], \label{p10genemod}
\end{equation}
Following the same path as above, in the limit of infinite viscosity, 
the translation velocity $V$ is found to be
\begin{equation}
V \simeq V_0 \left( 1 - \frac{5}{V_0} - \frac{5}{V_0^2} + \frac{30}{V_0^3} - \frac{30}{V_0^4}  \right)   \label{V2approxmod}
\end{equation}
when the effect of the boundary layer is taken into account.

\subsection{Melt production \label{SectionMeltProduction}}

We define the rate of melt production $\dot M$ as the volume of melt produced at the surface of the inner core by unit area and unit of time, averaged over the ICB. 
In the case of a pure translation, the volume of melt produced by unit of time is simply given by the translation velocity $V$ multiplied by  the cross-section $\pi r_\mathrm{ic}^2$ of the inner core, so $\dot M$ is given by
\begin{equation}
\dot M = \frac{V \times \pi r_\mathrm{ic}^2}{4 \pi r_\mathrm{ic}^2} = \frac{V}{4}.  \label{MeltingRate1}
\end{equation}

For a more general inner core flow, $\dot M$ can be calculated from the radial velocity at the ICB as
\begin{equation}
\dot M = \frac{1}{2} \overline{{|u_r(r_\mathrm{ic}) - \dot r_\mathrm{ic}|}} 	= \frac{1}{8 \pi} \int_{\theta,\phi} |u_r(r_\mathrm{ic}) - \dot r_\mathrm{ic}| \sin \theta\, d\theta d\phi, \label{DefMeltingRate}
\end{equation}
where the overbar $\overline{\dots}$ denotes the average over a spherical surface.
In the case of a $l=1$, $m=0$ flow, this reduces to 
\begin{equation}
\dot M = \frac{1}{4 \pi} \int_{\theta,\phi}  |p_1^0 \cos \theta| \sin \theta\, d\theta d\phi = \frac{1}{2} |p_1^0|.
\end{equation}
and gives $\dot M=V/4$ for a pure translation, which has $p_1^0=V/2$.

\section{Numerical results and scaling laws}
\label{Numerical_results}

\begin{figure*}
\begin{center}
\includegraphics[width=0.8\linewidth]{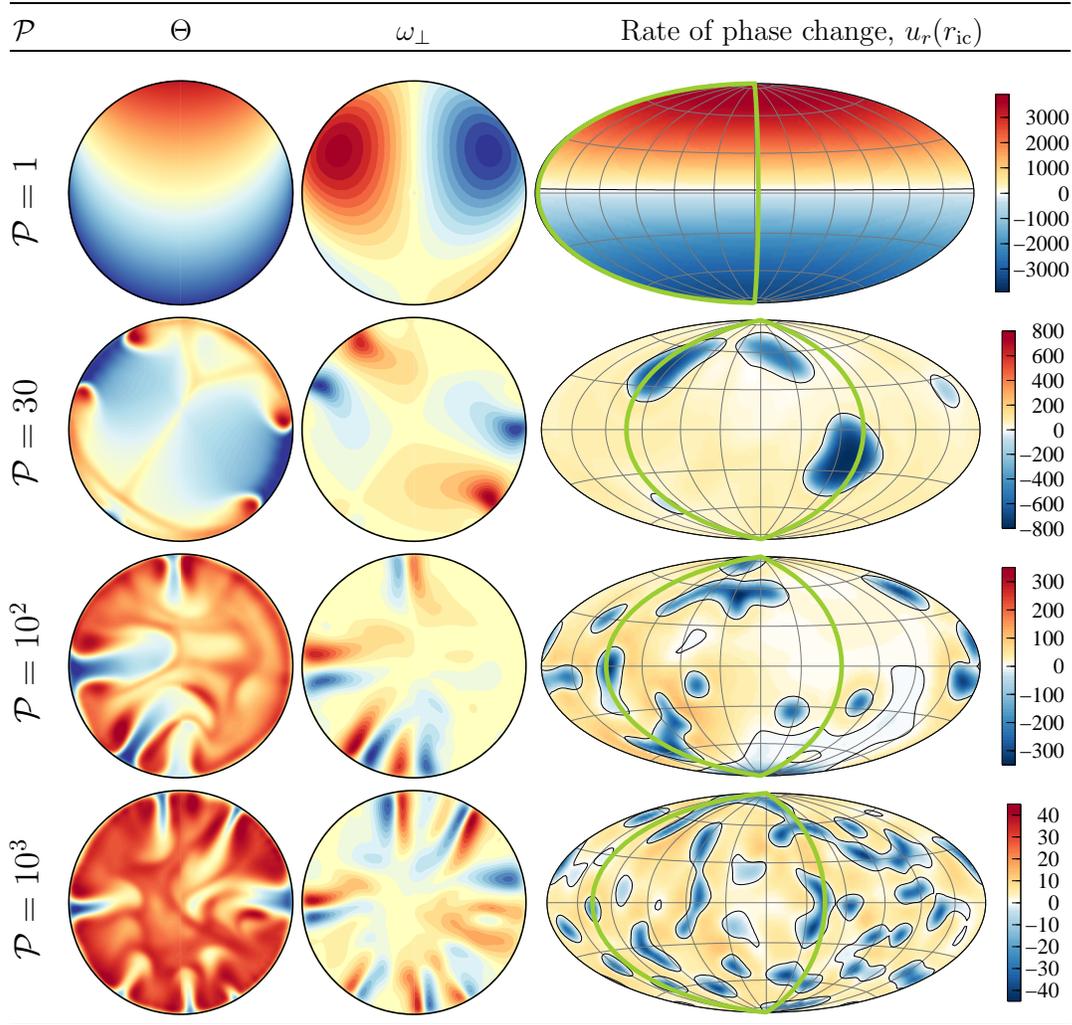}
\end{center}
\caption{Snapshots from numerical simulations with $Ra=10^7$ and $\mathcal{P}=1$, 30, 100, and $10^3$, showing potential temperature $\Theta$ (first column), azimuthal vorticity $\omega_\perp$ (second column), and radial velocity $u_r(r_\mathrm{ic})$ at the outer boundary (third column).
\label{Snapshots}
}
\end{figure*}

\subsection{Method}

The code is an extension of the one used in \cite{Deguen2011a}, with the boundary condition derived in section \ref{gov_eqns} now implemented.
The system of equations derived in section \ref{gov_eqns}  is solved in 3-D, using a spherical harmonic expansion for the horizontal dependence and a finite difference scheme in the radial direction. 
The radial grid can be refined below the ICB if needed.
The non-linear part of the advection term in the temperature equation is evaluated in the physical space at each time step. A semi-implicit Crank- Nickolson scheme is implemented for the time evolution of the linear terms and an Adams-Bashforth procedure is used for the non-linear advection term in the heat equation. 
The temperature field is initialized with a random noise covering the full spectrum.
We use up to 256 radial points and 128 spherical harmonics degree.
Care has been taken that the ICB thermal boundary layer, which can be very thin in the translation mode, is always well resolved. 

The code has the ability to take into account the growth of the inner core and the evolution of the internal heating rate $S(t)$, which is calculated from the thermal evolution of the outer core \citep{Deguen2011a}.
In this section, we will first focus on simulations with a constant inner core radius and steady thermal forcing (internal heating rate $S$ constant). 
Simulations with an evolving inner core will be presented in section \ref{application}.

Each numerical simulation was run for at least 10 overturn times $r_\mathrm{ic}/U_\mathrm{rms}$, where $U_\mathrm{rms}$ is the RMS velocity in the inner core.

\subsection{Overview}

As already suggested by the linear stability analysis (section \ref{linear_stability}) and the small $\mathcal{P}$ analytical model (section \ref{Small_P}), the translation mode is expected to be dominant when  $\mathcal{P}$ is small.
This is confirmed by our numerical simulations. 
As an example, figure \ref{Snapshots} shows outputs of simulations with the same Rayleigh number value of $Ra=10^7$ and $\mathcal{P}=1$, 30, $10^2$, and $10^3$.
Snapshots of cross-sections of the potential temperature field and vorticity (its component perpendicular to the cross-section plane) are shown in the first and second columns, and maps of radial velocity $u_r(r_\mathrm{ic})$ at the ICB are shown in the third column. $u_r(r_\mathrm{ic})$ is equal to the local phase change rate, with positive values corresponding to melting and negative values corresponding to solidification.

At the lowest $\mathcal{P}$ ($\mathcal{P}=1$), the translation mode is clearly dominant, with the pattern of temperature and vorticity similar to the predictions of the analytical models of section \ref{Small_P} shown in figure \ref{Analytical_T_Vorticity}.
In contrast, the convection regime at the largest $\mathcal{P}$ ($\mathcal{P}=10^3$) appears to be qualitatively similar to the regime observed with impermeable boundary conditions \citep{Weber92,Deguen2011a}, which corresponds to the limit $\mathcal{P}\rightarrow \infty$.
At  the Rayleigh number considered here, convection is chaotic and takes the form of cold plumes originating from a thin thermal boundary layer below the ICB, with a passive upward return flow.
At intermediate values of $\mathcal{P}$ ($\mathcal{P}=30$ and $10^2$), phase change has still a significant effect on the pattern of the flow, with large scale components of the flow enhanced by phase change at the ICB, in qualitative agreement with the prediction of the linear stability analysis.
Note that at $\mathcal{P}=10^2$, there is still a clear hemispherical pattern, with plumes originating preferentially from one hemisphere.

\begin{figure}
\begin{center}
\includegraphics[width=\columnwidth,clip=true]{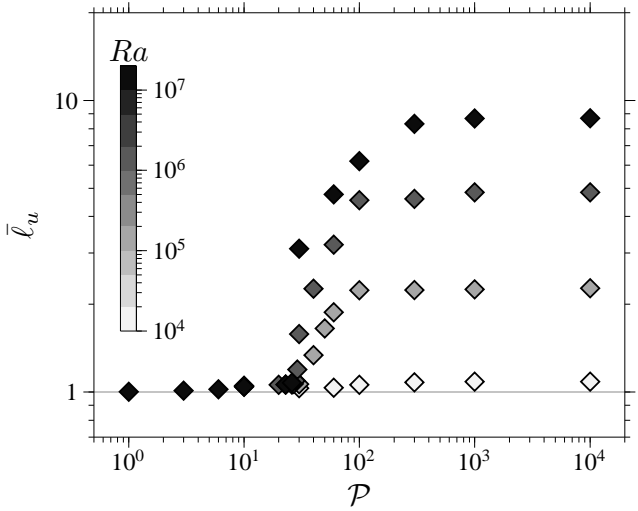}
\end{center}
\caption{Mean degree $\bar \ell_u $ of the kinetic energy (as defined in Equation \ref{MeanDegree}), as a function of $\mathcal{P}$, for simulations with $Ra=10^4$, $10^5$, $10^6$, and $10^7$. The grey scale of the markers give the Rayleigh number of the simulation. $\bar \ell_u $ is close to 1 for $\mathcal{P}\lesssim 29$ for all $Ra$, although the departure from 1 increases with $Ra$ when $\mathcal{P}$ approaches 29 from below.
\label{l_mean}
}
\end{figure}

More quantitative informations on the structure of convection can be found by estimating a characteristic length scale of the flow. 
We calculate here the mean degree $\bar \ell_u$ of the flow from the time averaged kinetic energy spectrum, defined by \cite{Christensen2006} as
\begin{equation}
\bar \ell_u = \frac{\sum_\ell \ell E_k^\ell}{E_k}, \label{MeanDegree}
\end{equation}
where
\begin{equation}
E_k^\ell = \frac{1}{2} \sum_m (u_\ell^m)^{2}\ \text{and}\ E_k = \sum_\ell E_k^\ell.
\end{equation}
With this definition, $\bar \ell_u\rightarrow1$ if the flow is dominated by degree 1 components, as in the translation mode, and increases as the characteristic lengthscale of the flow decreases.

Figure \ref{l_mean} shows the calculated value of $\bar \ell$ for $Ra=10^4$, $10^5$, $10^6$ and $10^7$ as a function of $\mathcal{P}$.
$\bar \ell$ remains very close to 1 as long as  $\mathcal{P}$ is smaller than a transitional value $\mathcal{P}_t\simeq 29$. 
There is a rapid increase of $\bar \ell_u$ above $\mathcal{P}_t$, showing the emergence of smaller scale convective modes at the transition between the translation mode and the high-$\mathcal{P}$ regime.
We interpret this sharp transition as being due to  the negative feedback that the secondary flow and smaller scale convection have on the translation mode : advection of the potential temperature field by the secondary flow decreases the strength of its degree one component and therefore weakens the translation mode, which in turn give more time for smaller scale convection to develop, weakening further the degree one heterogeneity.   
The value of $\mathcal{P}_t$ does not seem to depend on $Ra$ in the range explored here.
Figure \ref{l_mean} further shows that ICB phase change has a strong influence on the flow up to $\mathcal{P}\simeq 300$, which is confirmed by direct visualization of the flow structure.

\begin{figure}
\begin{center}
\includegraphics[width=\columnwidth,clip=true]{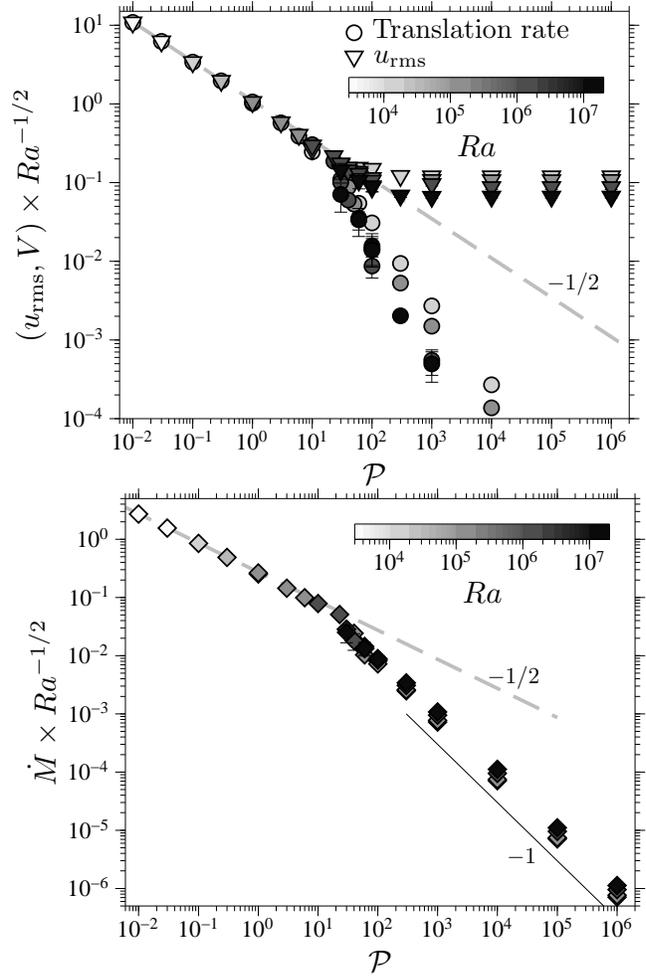}
\end{center}
\caption{\textbf{a)} RMS velocity (triangles) and translation velocity (circles) as a function of $\mathcal{P}$, for Rayleigh numbers between $3 \times 10^{3}$ and $10^7$ (grey scale). 
The inner core translation rate is found by first calculating the net translation rate $V_{i=x,y,z}$ of the inner core in the directions $x,y,z$ of a cartesian frame, given by the average over the volume of the inner core of the velocity component $u_{i=x,y,z}$ (which can be written as functions of the degree 1 components of the poloidal scalar at the ICB, see equation \ref{TranslationVelocityPoloidal} in Appendix \ref{Appendix_Tr}).
We then write the global translation velocity as $V=\sqrt{V_x^2+V_y^2+V_z^2}$.
The grey dashed  line shows the prediction of the rigid inner core model.
\textbf{b)} $\dot M\times Ra^{-1/2}$ as a function of $\mathcal{P}$, the grey scale of the markers giving the value of $Ra$.
The grey dashed  line shows the prediction of the rigid inner core model, showing excellent agreement between the theory and the numerical calculations for $\mathcal{P}$ small. 
\label{RMS_Tr_Melting}
}
\end{figure}

Figure \ref{RMS_Tr_Melting}a shows the translation rate $V$ (circles) and time averaged RMS velocity (triangles) as a function of $\mathcal{P}$ for various values of $Ra$. 
Here both $V$ and $U_\mathrm{rms}$ are multiplied by $Ra^{-1/2}$. 
The grey dashed line shows the analytical prediction for the translation rate in the rigid inner core limit.
Below $\mathcal{P}_t$, there is a good quantitative agreement between the numerical results and the analytical model.
The fact that $U_\mathrm{rms}\simeq V$ for $\mathcal{P}<\mathcal{P}_t$ indicates that there is, as expected, negligible deformation in this regime.
$V$ and $U_\mathrm{rms}$ diverge for $\mathcal{P}>\mathcal{P}_t$, the translation rate becoming rapidly much smaller than the RMS velocity. 
As already suggested by the evolution of $\bar \ell$, phase change at the ICB has still an effect on the convection for $\mathcal{P}$ up to $\sim 300$.
Phase change at the ICB has a positive feedback on the vigor of the convection:
melting occurs preferentially above upwelling, where the dynamic topography is positive, which enhances upward motion. 
Conversely, solidification occurs preferentially above downwellings, thus enhancing downward motions.
This effect becomes increasingly small as $\mathcal{P}$ is increased, and the RMS velocity reaches a plateau when $\mathcal{P}\gtrsim 10^3$, at which the effect of phase change at the ICB on the internal dynamics becomes negligible.

Figure \ref{RMS_Tr_Melting}b shows the rate of melt production (defined in Equation \eqref{DefMeltingRate}), multiplied by $Ra^{-1/2}$, as a function of $\mathcal{P}$ for various values of $Ra$. 
Again, the prediction of the rigid inner core model (Equation \eqref{MeltingRate1}, grey dashed line in Figure \ref{RMS_Tr_Melting}b) is in very good agreement with the numerical results as long as $\mathcal{P}<\mathcal{P}_t$.
For $\mathcal{P}>\mathcal{P}_t$, the rate of melt production appears to be inversally proportional to $\mathcal{P}$.

\subsection{Scaling of translation rate, convective velocity, and melt production}

We now turn to a more quantitative description of the small-$\mathcal{P}$ and large-$\mathcal{P}$ regimes. We first compare the results of numerical simulations at $\mathcal{P}<\mathcal{P}_t$ with the analytical models developed in section \ref{Small_P}. We then focus on the large-$\mathcal{P}$ regime, and develop a scaling theory for inner core thermal convection in this regime,  including a scaling law for the rate of melt production.

\subsubsection{Translation mode}

\begin{figure}
\begin{center}
\includegraphics[width=\columnwidth]{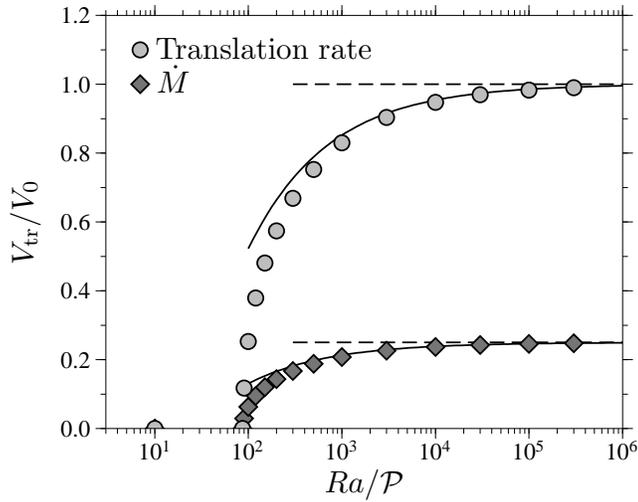}
\end{center}
\caption{
Translation rate and melt production, normalized by the low $\mathcal{P}$ limit estimate given by Equation \eqref{V2approx}, as a function of $Ra/\mathcal{P}$, for $\mathcal{P}=10^{-2}$.
\label{RaP_V}
}
\end{figure}

Figure \ref{RaP_V} shows the translation rate (circles) and the rate of melt production $\dot M$ (diamonds), normalized by the rigid inner core estimate given by Equation \eqref{V2approx}, as a function of $Ra/\mathcal{P}$, for $\mathcal{P}=10^{-2}$.
The translation rate increases from zero when $Ra/\mathcal{P}$ is higher than a critical value $(Ra/\mathcal{P})_c$ which is found to be  in excellent agreement with the prediction of the linear stability analysis.
Increasing $Ra/\mathcal{P}$ above $(Ra/\mathcal{P})_c$, the translation rate increases before asymptoting toward the prediction of the rigid inner core model (dashed line).
The prediction of our model including a boundary layer correction (Equation \eqref{V2approxmod}, black line in Figure \ref{RaP_V}) is in good agreement with the numerical results for $Ra/\mathcal{P}\gtrsim10^3$, demonstrating that our analytical model captures fairly well the effect of the thermal boundary layer.
As expected (see section \ref{SectionMeltProduction}), the rate of melt production is equal to $1/4$ of the translation rate.

\begin{figure}
\begin{center}
\includegraphics[width=\columnwidth,clip=true]{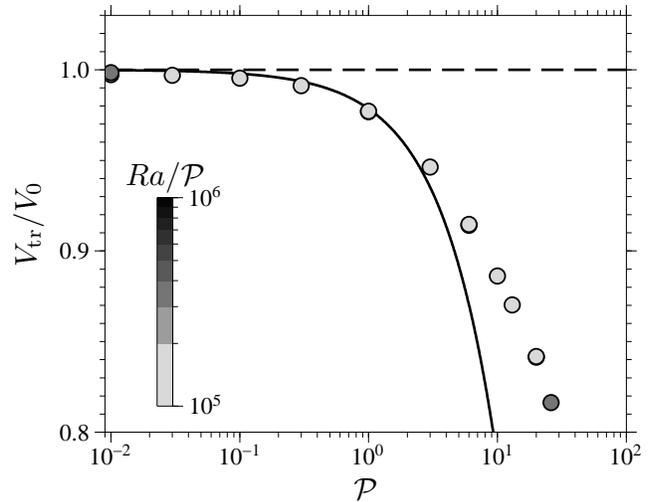}
\end{center}
\caption{Translation rate (normalized by the low $\mathcal{P}$ limit estimate given by Equation \eqref{V2approx}) as a function of   $\mathcal{P}$, for different values of $Ra/\mathcal{P}$.
The thick black line show the prediction of the $\mathcal{O}(\mathcal{P})$ model given by equation \eqref{TranslationOrdreP}.
\label{EffectOfP}
}
\end{figure}

Figure \ref{EffectOfP} shows the effect of increasing $\mathcal{P}$ on the translation rate.
In this figure, we have kept only simulations with $Ra/\mathcal{P}$ larger than $10^5$ to minimize the effect of the boundary layer, and further corrected the translation velocity with the boundary layer correction (Equation \eqref{V2approxmod}) found in section \ref{SectionBoundaryLayer}, in order to isolate as much as possible the effect of $\cal{P}$ on the translation mode.
The $\mathcal{O}(\mathcal{P})$ model developed in section \ref{SectionTranslationOrdreP} (Equation \eqref{TranslationOrdreP}, black line) agrees with the numerical simulations within 1\% for $\mathcal{P}$ up to $\sim 3$, but fails to explain the outputs of the numerical simulations when $\mathcal{P}$ is larger, which indicates that higher order terms in $\mathcal{P}$ become important.

Overall, our analytical results (stability analysis and finite amplitude models) are in very good agreement with our numerical simulations when $\mathcal{P}$ is small, which gives support to both our theory and to the validity of the numerical code.

\subsubsection{Plume convection}
\label{PlumeConvection}

\begin{figure}
\begin{center}
\includegraphics[width=\columnwidth,clip=true]{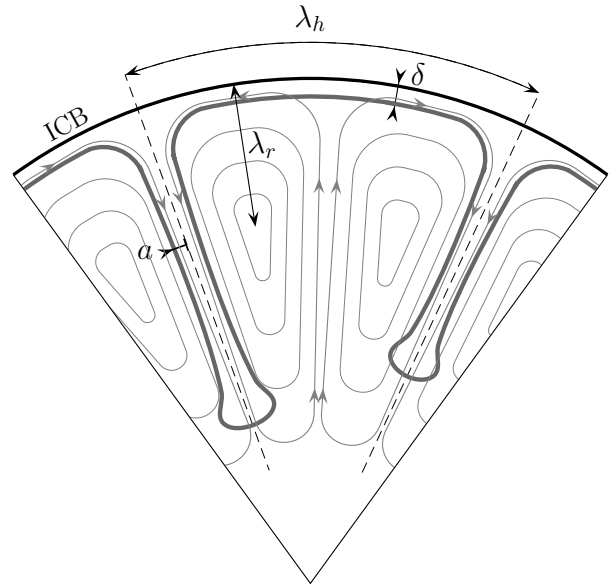}
\end{center}
\caption{A schematic of inner core plume convection, and definition of the length scales used in the scaling analysis.
Streamlines of the flow are shown with thin arrowed grey lines.  \label{ConvectionSchematic}
}
\end{figure}

If $\mathcal{P}$ is large, the translation rate of the inner core becomes vanishingly small, but, as long as $\mathcal{P}$ is finite, there is still a finite rate of melt production associated with the smaller scale topography arising from plume convection.
A scaling for the melt production in the limit of large $\mathcal{P}$ and large $Ra$ can be derived from scaling relationship  for infinite Prandtl number convection with impermeable boundaries.
\cite{Parmentier2000} derived a set of scaling laws for high Rayleigh number internally heated thermal convection in a cartesian box, in the limit of infinite Prandtl number, but we found significant deviations from their model in our numerical simulations, which we ascribe to geometrical effects due to the spherical geometry.
We therefore propose a set of new scaling laws for convection in a full sphere with internal heating.

Quantities of interest are the horizontal and vertical velocities $u$ and $w$, the mean inner core potential temperature $\left< \Theta \right>$, the thermal boundary layer thickness $\delta$, the thermal radius of the plumes $a$, the average plume spacing $\lambda_h$, and a length scale for radial variations of the velocity, which we note $\lambda_r$ (see Figure \ref{ConvectionSchematic}).
The horizontal length scale $\lambda_h$ is related to the number $N$ of plumes per unit area by $N\sim1/\lambda_h^2$.

\begin{figure*}
\begin{center}
\includegraphics[width=0.8\linewidth,clip=true]{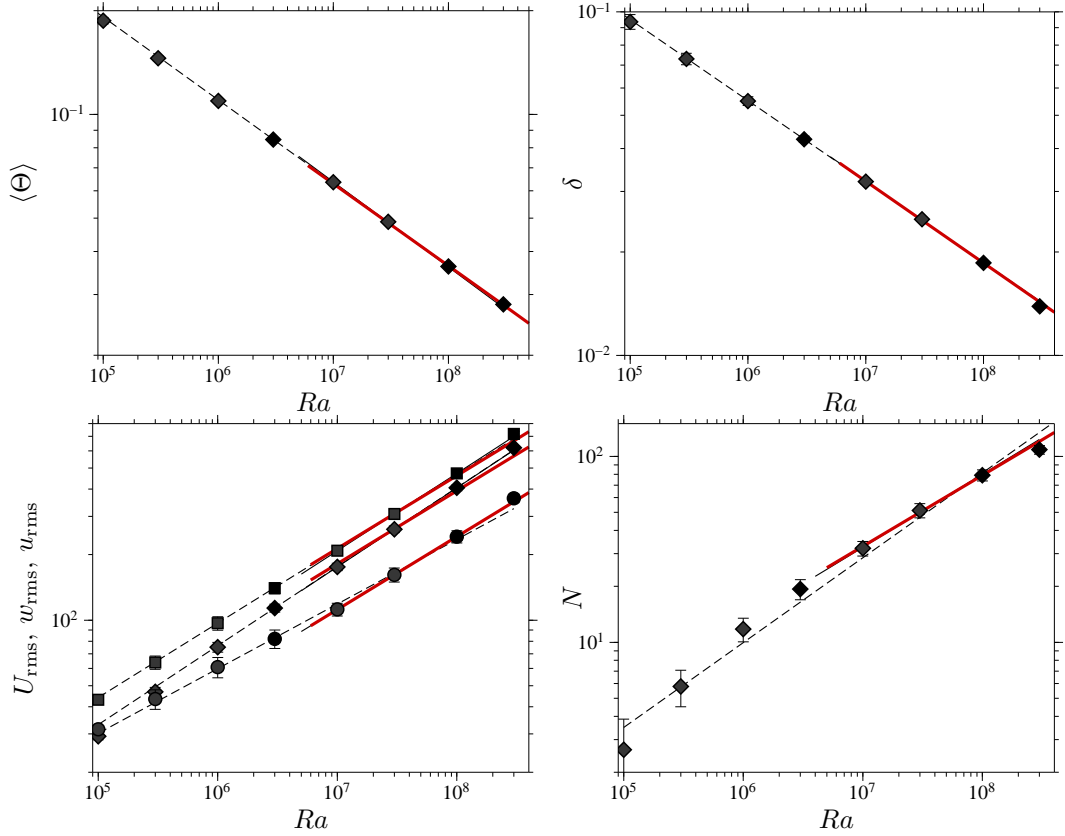}
\end{center}
\caption{\textbf{a)} Mean potential temperature $\left< \Theta \right>$ as a function of $Ra$ for $\mathcal{P}$ larger than $10^3$. 
\textbf{b)} Boundary layer thickness $\delta$.
\textbf{c)} RMS velocity $U_\mathrm{rms}$ (squares), RMS vertical velocity $w_\mathrm{rms}$ (diamonds), and RMS horizontal velocity $u_\mathrm{rms}$ (circles). 
\textbf{d)} Number of plumes $N$ per unit surface. 
In figures \textbf{a)} to \textbf{d)}, the thick red lines show the predictions of the scaling theory developed in section \ref{PlumeConvection} with $\beta=-0.238$.
The dashed lines show the result of the individual least square inversion for each quantity for $Ra\geq 10^5$.
\label{ScalingConvection}
}
\end{figure*}

Outputs of numerical simulations ($\left< \Theta \right>$, $\delta$, RMS velocity $U_\mathrm{rms}$, RMS radial velocity $w_\mathrm{rms}$, RMS horizontal velocity $u_\mathrm{rms}$, $N$) are shown in Fig.\ref{ScalingConvection}a-d for $Ra$ between $10^5$ and $3\times 10^8$.
The boundary layer thickness $\delta$ is estimated as the ratio of the mean potential temperature in the inner core, $\left< \Theta \right>$, over the time and space averaged potential temperature gradient at the ICB : $\delta = - \left< \Theta \right>/ \left< \partial \Theta/\partial r\right>_\mathrm{icb}$. 
The time-averaged number $N$ of plume per unit area is estimated by counting plumes on horizontal surfaces on typically $50$ different snapshots. 
Both $\left< \Theta \right>$ and $\delta$ follow well-defined power law behaviors over this range of $Ra$. 
In contrast, the RMS velocities and plume density $N$ seem to indicate a change of behavior at $Ra$ close to $10^7$.
For $Ra< 10^7$, the vertical velocity increases faster than the horizontal velocity, while at larger $Ra$ horizontal and vertical velocities increase with $Ra$ at roughly the same rate.

We start our analysis by first noting that under statistically steady state conditions, the heat flux at the ICB must be equal, in a time-averaged sense, to the heat production within the inner core. 
In the thermal boundary layer, heat transport is dominated by conduction and the non-dimensional heat flux $-\left< \partial \Theta/\partial r\right>_\mathrm{icb}$ is equal to $\left< \Theta \right>/\delta$. This must be in balance with the non-dimensional internal heat production.  
With our scaling, the mean potential temperature gradient should be equal to -2 on average, which implies that $\delta$ should be equal to $\left< \Theta \right>/2$.
We can therefore write
\begin{equation}
\delta  = \frac{\left< \Theta \right>}{2} \sim  Ra^\beta,
\end{equation}
where $\beta$ is to be determined.
We further assume that the thickness of the thermal boundary layer is set by a local stability criterion, \textit{i.e.} that the boundary layer Rayleigh number $Ra_\delta=(\alpha \rho_s g_\mathrm{icb} \Theta \delta^3)/(\kappa \eta)$ is on average equal to some constant, which is equivalent to state that $Ra_\delta \sim 1$.
Using non-dimensional $\left< \Theta \right>$ and $\delta$, $Ra_\delta$ is related to the inner core Rayleigh number $Ra$ by $Ra_\delta = Ra\, \left< \Theta \right> \delta^3$. 
Given that $\left< \Theta \right> \sim \delta$, this implies that  $Ra \left< \Theta \right> \delta^3 \sim Ra\, \delta^4 \sim 1$, which gives $\beta=-1/4$.

The best fit of the numerical results (Figure \ref{ScalingConvection}a and b) gives $\left< \Theta \right> \sim Ra^{-0.240\pm 0.005}$ and $\delta \sim Ra^{-0.236 \pm 0.003}$, in fair agreement with the predicted scaling.
In cartesian geometry, \cite{Parmentier2000} found $\beta=-0.2448$.
\cite{Deschamps2012} found $\beta=-0.238$ for thermal convection in internally heated spherical shells.

The vertical plume velocity $w$ is set by a balance between the buoyancy stress, $\sim Ra\left< \Theta \right> a$, and the viscous stress, $\sim w/\lambda_h$. 
This gives
\begin{equation}
\frac{w}{\lambda_h} \sim Ra\,  \left< \Theta \right>\, a. \label{ScalingSystem1}
\end{equation}
In addition, the heat flux advected by the plumes, $N w  \left< \Theta \right> a^2$, must scale as the ICB heat flux, which, as already discussed above, must be $\sim 1$.
Since the number of plumes  per unit area is $N\sim 1/\lambda_h^2$, this gives
\begin{equation}
1 \sim \frac{a^2}{\lambda_h^2} \left< \Theta \right> w. \label{ScalingSystem2}
\end{equation}
The plume thermal radius $a$ is related to the thermal boundary layer thickness through the conservation of mass, which when applied at the roots of the plumes implies that
\begin{equation}
\delta u \sim a w.   \label{ScalingSystem3}
\end{equation}
Finally, conservation of mass in one convective cell implies that
\begin{equation}
\frac{u}{\lambda_h} \sim \frac{w}{\lambda_r}. \label{ScalingSystem4}
\end{equation}

This gives four equations (\ref{ScalingSystem1}-\ref{ScalingSystem4}) for five unknowns ($u$, $w$, $a$, $\lambda_h$, $\lambda_r$). 
The system can be solved if additional assumptions are made on the scaling of $\lambda_r$.
For high $Pr$, low $Re$ convection, a natural choice would be to assume that  radial variations of $w$ occur at the scale of the radius of the inner core. 
This implies $\lambda_r\sim 1$, and solving the system of equations (\ref{ScalingSystem1}-\ref{ScalingSystem4}) with $\beta=-0.24$ gives $a\sim Ra^{-0.14}$, $u\sim Ra^{0.82}$, $w\sim Ra^{0.72}$ and $N\sim Ra^{-0.2}$, which agrees very poorly with the numerical results. 

This poor agreement might be due to the spherical geometry. In a sphere, plumes converge toward each others while sinking, which is not the case in cartesian boxes, and is not a very significant effect in a spherical shell for which, like in Earth's mantle, the radius of the inner shell is a significant fraction of that of the outer shell.
If $Ra$ is large and the average plume spacing is small compared to the inner core radius, we might expect that the geometry of the convective cells becomes self-similar, with $\lambda_r\sim \lambda_h$.
With this assumption, we obtain
\begin{align}
\lambda_h \sim \lambda_r &\sim Ra^{1+5\beta},\label{ScalingBeta1}\\
u\sim w &\sim Ra^{2+7\beta},  \label{ScalingBeta2} \\
a \sim \delta \sim \left< \Theta \right> &\sim Ra^\beta. \label{ScalingBeta3}
\end{align}

Assuming a scaling of the form given by equations (\ref{ScalingBeta1}-\ref{ScalingBeta3}), it is possible to inverse simultaneously all variables for $\beta$, the result of the inversion being $\beta=-0.238\pm0.003$ ($\pm1\,\sigma$). 
The prediction of equations (\ref{ScalingBeta1}-\ref{ScalingBeta3}) with this value of $\beta$ are shown with red  lines in Fig.\ref{ScalingConvection}a-d for $Ra\geq 10^7$. 
They agree with the numerical outputs almost as well as individual inversions, which demonstrates the self-consistency of our scaling theory.

We can now derive a scaling for the rate of melt production $\dot M$.
The starting point  is the continuity of the normal stress at the ICB, given by Equation \ref{normalstressfa}.
The local melting/solidification rate is given by the value of  $u_r-\dot r_\mathrm{ic}$ at the ICB ($u_r-\dot r_\mathrm{ic}>0$ means melting, and $u_r-\dot r_\mathrm{ic}<0$ means solidification)  which, according to Equation \ref{normalstressfa}, can be written as 
\begin{equation}
u_r-\dot r_\mathrm{ic} = \mathcal{P}^{-1} \left( -2 \frac{\partial u_r}{\partial r} + \hat p  \right).
\end{equation}
As discussed above, we have
\begin{equation}
\left.\frac{\partial u_r}{\partial r}\right|_\mathrm{icb} \sim \frac{w}{\lambda_r} \sim Ra^{1+2\beta}.
\end{equation}
The dynamic pressure is given by the horizontal component of the Stokes equation,
\begin{equation}
0 = -\nabla_H \hat p + (\Delta u)_H
\end{equation}
which implies that
\begin{equation}
\hat p \sim \frac{u}{\lambda_h} \sim Ra^{1+2\beta}.
\end{equation}
Both terms follow  the same scaling,
which implies that the global rate of melt production scales as
\begin{equation}
\dot M \sim Ra^{1+2\beta} \mathcal{P}^{-1}. \label{MeltScaling}
\end{equation}
With $\beta=-0.238\pm0.003$, we obtain $\dot M \sim Ra^{0.524\pm0.006} \mathcal{P}^{-1}$.

\begin{figure}
\begin{center}
\includegraphics[width=\columnwidth,clip=true]{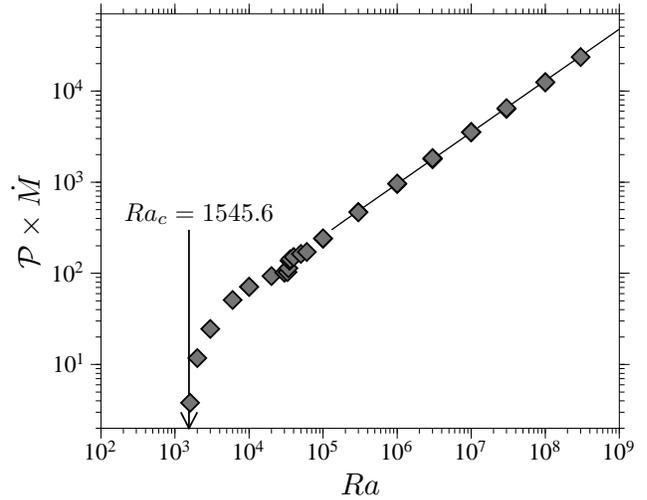}
\end{center}
\caption{Rate of melt production (multiplied by $\mathcal{P}$) as a function of $Ra$, for numerical simulations with $\mathcal{P} \geq 10^3$. 
The value of the critical Rayleigh number as predicted by the linear stability analysis in the limit of infinite $\mathcal{P}$ (Equation \ref{Rac_largeP}, $Ra_c=1545.6$) is indicated by the arrow.
\label{MeltingRate}
}
\end{figure}

Figure \ref{MeltingRate}b shows $\mathcal{P} \dot M$ as a function of $Ra$, for $\mathcal{P}\geq10^3$ corresponding to the plume convection regime.
There is an almost perfect collapse of the data points, which supports the fact that $\dot M \propto \mathcal{P}$ in this regime.
The kink in the curve at $Ra\simeq 3\times10^4$ corresponds to the transition from steady convection to unsteady convection.
Above this transition, the data points are well fitted by a power-law of the form $\dot M =a \mathcal{P}^{-1} Ra^{b}$. Least square regression for $Ra\geq 3\times 10^5$ gives $a= 0.46\pm0.04$ and $b=0.554\pm0.006$, in reasonable agreement with the  value found above.
In dimensional terms, $\dot M \simeq a ({\kappa}/{r_\mathrm{ic}}) \mathcal{P}^{-1}Ra^{b}$ and the mass flux of molten material is $\rho_\mathrm{ic} \dot M \simeq a {k}/({c_p r_\mathrm{ic}}) \mathcal{P}^{-1}Ra^{b}$.

\section{Application}
\label{application}

\begin{figure*}
\includegraphics[width=\linewidth]{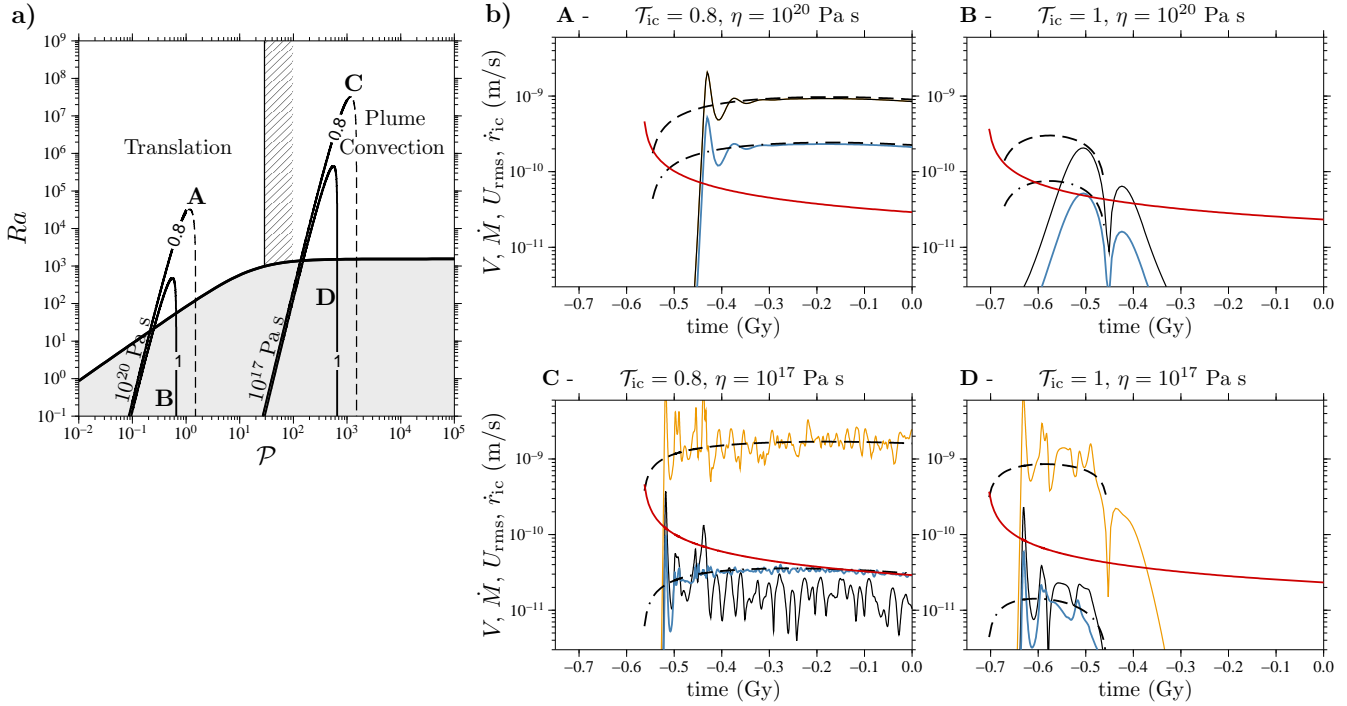}
\caption{
\textbf{a)} Trajectories of the inner core state in a $Ra-\mathcal{P}$ space, for the four cases \textbf{A}, \textbf{B}, \textbf{C}, and \textbf{D}  discussed in the text.
The line annotations give the value of $\mathcal{T}_\mathrm{ic}$ for each case.
The dashed lines shows the future trajectory of the inner core.
\textbf{b)}
Time evolution of $V$, $\dot M$, $U_\mathrm{rms}$ and $\dot r_\mathrm{ic}$ for cases \textbf{A} to \textbf{D}.
Red line~: inner core growth rate $\dot r_\mathrm{ic}$.
Black line~: translation rate $V$.
Orange line~: RMS velocity $U_\mathrm{rms}$.
Blue line~: dimensional melting rate $(\kappa/r_\mathrm{ic})\dot M$.
Predictions for the RMS velocity (or translation velocity in the translation regime) and melting rate $\dot M$ are shown with thick dashed and dash-dotted lines, respectively. 
In the $\eta=10^{20}$ Pa.s cases, the translation model (Equation \ref{V2approxmod}) is used to predict $V$ and $\dot M$.
In the $\eta=10^{17}$ Pa.s case, the high-$\mathcal{P}$ scaling is used for $U_\mathrm{rms}$ and $\dot M$.
In the $\eta=10^{20}$ Pa.s, $\mathcal{T}_\mathrm{ic}=0.8$ and $\mathcal{T}_\mathrm{ic}=1$ cases, the translation rate and the RMS velocity are equal.
For these simulations, the Rayleigh number was calculated assuming a thermal conductivity $k=79$ W.m$^{-1}$.K$^{-1}$ and a phase change timescale $\tau_\phi=1000$ yr.
Values of other physical parameters used for these runs are summarized in Table \ref{parameters}.
\label{EvolutiveSimulations}
}
\end{figure*}

\subsection{Evolutive models}

The analytical model for the translation mode and the scaling laws for large-$\mathcal{P}$ convection derived in the previous sections strictly apply only to convection with $r_\mathrm{ic}$ and $S$ constant.
We therefore first check that our models correctly describe inner core convection when $r_\mathrm{ic}$ and $S$ are time-dependent, by comparing their predictions with the outcome of numerical simulations with inner core growth and thermal history determined from the core energy balance.

To account for the inner core secular evolution, we follow the procedure explained in \cite{Deguen2011a}, where the growth of the inner core and its cooling rate are determined from the core energy balance.
In this framework, a convenient way to write $S(t)$ is
\begin{equation}
S =   \frac{\rho_s g' \gamma T}{K_S} 3 \kappa \left[ f(r_\mathrm{ic})\,\mathcal{T}_\mathrm{ic}^{-1} - 1  \right], \label{S_final_2}
\end{equation}
where $f(r_\mathrm{ic})$ is a decreasing order one function of $r_\mathrm{ic}$ defined in \cite{Deguen2011a} (Equation 19, page 1104), $g'=dg/dr$, $K_S$ is the isentropic bulk modulus, $\gamma$ the Gr\"uneisen parameter, and
\begin{equation}
\mathcal{T}_\mathrm{ic}=\left( \frac{dT_s}{dT_\mathrm{ad}}-1 \right)^{-1} \frac{\tau_\mathrm{ic}}{\tau_\kappa}, \label{Tau_ic}
\end{equation}
where $\tau_\mathrm{ic}$ is the age of the inner core, $\tau_\kappa=r_\mathrm{ic}^{*2}/(6\kappa)$ is the current inner core thermal diffusion time, and $dT_s/dT_\mathrm{ad}$ is  the ratio of the Clapeyron slope $dT_s/dP$ to the adiabat $dT_\mathrm{ad}/dP$. 
The non-dimensional inner core age $\mathcal{T}_\mathrm{ic}$ is a convenient indicator of the thermal state of the inner core, with $\mathcal{T}_\mathrm{ic}<1$ implying unstable stratification for most of inner core history (see \cite{Deguen2011a}, Figure 3a).
We give for reference in table \ref{Table_Tau_ic} the  values of the age of the inner core $\tau_\mathrm{ic}$  corresponding to $\mathcal{T}_\mathrm{ic}=0.8$, 1, and 1.2, for a thermal conductivity equal to 36, 79, and 150 W.m$^{-1}$.K$^{-1}$.
With the inner core growth history determined from the core energy balance and $S(t)$ calculated from Equation \eqref{S_final_2}, the evolution of $Ra(t)$ and $\mathcal{P}(t)$ can then be calculated.

Figure \ref{EvolutiveSimulations}a shows the trajectories of the inner core state in a $Ra-\mathcal{P}$ space for four different scenarios, superimposed on a regime diagram for inner core thermal convection.
According to Equation \eqref{tauphi}, the ICB phase change timescale  scales as $\tau_\phi \propto r_\mathrm{ic}^{-1}$, and therefore $\mathcal{P}\propto r_\mathrm{ic}$ always increases  during inner core history.
In contrast, the evolution of $Ra(t)$ is non monotonic, with the effect of the increasing inner core radius and gravity opposing the decrease with time of the effective heating rate $S(t)$.
Because $S$  eventually becomes negative at some time in inner core history, $Ra$ reaches a maximum  before decreasing and eventually becoming negative, 
resulting in a bell shaped trajectory of the inner core in the $Ra-\mathcal{P}$ space.
The maximum in $Ra$ may or may not have been reached yet, depending on the value of $\mathcal{T}_\mathrm{ic}$.

The scenarios \textbf{A}-\textbf{D} shown in Figure \ref{EvolutiveSimulations}a have been chosen to illustrate four different possible dynamic histories of the inner core.
In cases \textbf{A} and \textbf{C}, which have $\mathcal{T}_\mathrm{ic}=0.8$, $Ra$ remains positive and supercritical up to today, thus always permitting thermal convection.   
In cases \textbf{B} and \textbf{D}, which have $\mathcal{T}_\mathrm{ic}=1$, $Ra$ has reached a maximum early in inner core history, before decreasing below supercriticality, at which point convection is expected to stop.
In these two cases, only an early convective episode is expected \citep{Buffett2009,Deguen2011a}.
In cases \textbf{A} and \textbf{B}, which have $\eta=10^{20}$ Pa~s,   $\mathcal{P}(t)$ is always smaller than the transitional $\mathcal{P}_t$ and thermal convection therefore should be in the translation regime; Cases \textbf{C} and \textbf{D}, which have $\eta=10^{17}$ Pa~s, have $\mathcal{P}(t)>10^2>\mathcal{P}_t$ and thermal convection should be in the plume regime.

Figure \ref{EvolutiveSimulations}b shows  outputs from numerical simulations corresponding to the inner core histories shown in Figure \ref{EvolutiveSimulations}a.
The numerical results are compared to the predictions for the RMS velocity $U_\mathrm{rms}$ (equal to the translation rate $V_{tr}$ in the translation regime) and melting rate $\dot M$ from the analytical translation model (Equation \eqref{MeltingRate1}) and the large-$\mathcal{P}$ scaling laws (Equation \eqref{MeltScaling}).
The agreement is good in both the translation and plume convection regimes, except at the times of initiation and cessation of convection.

There is always a lag between when conditions become supercritical and when the amplitude of  convective motions become significant, due to the finite growth rate of the instability.
From Equation \eqref{GrowthRates}, the timescale for instability growth, $\tau=1/\sigma$, is approximately (in dimensional form)
\begin{equation}
\tau \simeq 5 \frac{r_\mathrm{ic}^2}{\kappa} \left( \frac{Ra}{\mathcal{P}} - \left.\frac{Ra}{\mathcal{P}}\right|_c  \right)^{-1} 
\simeq \left( \frac{r_\mathrm{ic}}{600\ \text{km}} \right)^2  \frac{90\ \text{My}}{\dfrac{Ra}{\mathcal{P}}\left/\dfrac{Ra}{\mathcal{P}}\right|_c - 1}
\end{equation}
in the translation regime, and
\begin{equation}
\tau \simeq 77 \frac{r_\mathrm{ic}^2}{\kappa} \left(  Ra  - Ra_c  \right)^{-1}
\simeq \left( \frac{r_\mathrm{ic}}{600\ \text{km}} \right)^2  \frac{80\ \text{My}}{{Ra}/{Ra}_c - 1}
\end{equation}
in the large-$\mathcal{P}$ regime.
In both cases, the timescale for the growth of the instability will typically be a few tens of My, thus explaining the delayed initiation of convection seen in the numerical simulations.

In cases \textbf{B} and \textbf{D}, the flow occurring after $t \simeq -0.46$ Gy, at a time where the models predict no motion (because $S<0$),  corresponds to a slow relaxation of the thermal heterogeneities left behind by the convective episode.

Apart during the initiation and cessation periods of convection, the models developed for steady internal heating and constant inner core radius agree very well with the full numerical calculations, and can therefore be used to predict the dynamic state of the inner core and key quantities including RMS velocity and melt production rate.

\begin{table}
\caption{Correspondence between $\tau_\mathrm{ic}$  and $\mathcal{T}_\mathrm{ic}$ for three values of inner core thermal conductivity, assuming $dT_s/dT_\mathrm{ad}=1.65\pm0.11$ \citep{Deguen2011a}.}
\label{Table_Tau_ic}
\begin{center}
\begin{tabular*}{\columnwidth}{@{}ll*{3}{c}} 
\toprule
	&					&		&	$k$ (W.m$^{-1}$.K$^{-1}$)	&	\\
	&					& 36		&	79	& 	150		\\
\midrule
						&	0.8	& $1.18\pm0.23$ Gy	& $0.54\pm0.11$ Gy	& $0.28\pm0.06$ Gy	\\
${\mathcal{T}_\mathrm{ic}}=$	& $1.0$ 	& $1.48\pm0.29$ Gy	& $0.68\pm0.13$ Gy	& $0.36\pm0.07$ Gy	\\
						& $1.2$	& $1.77\pm0.35$ Gy	& $0.81\pm0.16$ Gy	& $0.43\pm0.08$ Gy	\\
\bottomrule
\end{tabular*}
\end{center}
\end{table}

\subsection{Melt production}

Experiments by \citet{Alboussiere2010} have shown that the development of a stably stratified layer above the ICB by inner core melting is controlled by the ratio $\Phi_B$ of the buoyancy fluxes arising from the melting and freezing regions of the ICB.
By using the analytical translation model and the scaling laws for plume convection developed in the last two sections, we can now estimate today's value of $\Phi_B$ as a function of the state and physical properties of the inner core, and assess the likelihood of the origin of the F-layer by inner core melting.

With $\dot M$ being the non-dimensional rate of melt production defined in Equation \eqref{DefMeltingRate}, the mean solidification rate is  $(\kappa/r_\mathrm{ic})\dot M + \dot r_\mathrm{ic}$ from conservation of  mass.
The buoyancy flux associated with the release of dense fluid by melting can be written as $- \Delta \rho_\chi\, g_\mathrm{icb} (\kappa/r_\mathrm{ic}) \dot M$, while the buoyancy flux associated with the solidification is $\Delta \rho_\chi\, g_\mathrm{icb} [(\kappa/r_\mathrm{ic})\dot M + \dot r_\mathrm{ic}]$, where  $\Delta \rho_\chi$ is the fraction of the ICB density jump due to the compositional difference.
According to \cite{Alboussiere2010}'s experiments, a stratified layer is expected to form above the ICB if the magnitude of the buoyancy flux associated with melting is more than 80 \% of the buoyancy flux associated with solidification, \textit{i.e.} if
\begin{equation}
\Phi_B = \frac{ \Delta \rho_\chi\, g_\mathrm{icb} (\kappa/r_\mathrm{ic}) \dot M}{\Delta \rho_\chi\, g_\mathrm{icb} [(\kappa/r_\mathrm{ic})\dot M + \dot r_\mathrm{ic}]}= \frac{  \dot M}{\dot M + \dot r_\mathrm{ic}\, r_\mathrm{ic}/\kappa} > 0.8,
\end{equation}
which requires that
\begin{equation}
\frac{\kappa}{r_\mathrm{ic}}\dot M > 4\, \dot r_\mathrm{ic}.
\end{equation}
In the translation regime, in which $\dot M = V/4$, this requires that the rate of translation is at least 16 times larger than the mean solidification rate of the inner core.

\begin{figure}
\begin{center}
\includegraphics[width=\linewidth]{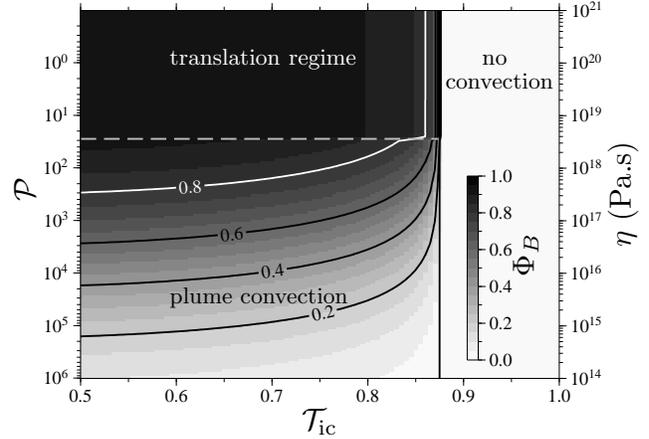}
\end{center}
\caption{Inner core regime diagram and map of the buoyancy ratio $\Phi_B$, as functions of $\mathcal{P}$ and $\mathcal{T}_\mathrm{ic}$.
The corresponding values of $\eta$ assuming $\tau_\phi=1000$ yr are given on the right hand size vertical axis.
According to \citet{Alboussiere2010}'s experiments, inner core melting can produce a stably stratified layer at the base of the outer core if $\Phi_B>0.8$ (white contour).
\label{MeltingRateRegime}
}
\end{figure}

The current inner core growth rate can be expressed as
\begin{equation}
\dot r_\mathrm{ic} = \frac{3\kappa}{r_\mathrm{ic}} \frac{f(r_\mathrm{ic}) }{\left(\frac{dT_s}{dT_\mathrm{ad}}-1\right)\mathcal{T}_\mathrm{ic}}
\end{equation}
where the function $f(r_\mathrm{ic})\simeq 0.8$ at the current inner core radius \citep{Deguen2011a}.
Using this expression, the buoyancy ratio $\Phi_B$ is 
\begin{equation}
\Phi_B = 1 - \left[ 1 +  \left(\frac{dT_s}{dT_\mathrm{ad}} -1 \right) \frac{\mathcal{T}_\mathrm{ic}}{3 f(r_\mathrm{ic})}  \frac{V}{4}  \right]^{-1}  	\label{PhiTranslation}
\end{equation}
in the translation regime, with the translation velocity $V$ given by Equation \eqref{V2approxmod}, and 
\begin{equation}
\Phi_B = 1 - \left[ 1 + \left(\frac{dT_s}{dT_\mathrm{ad}} -1 \right) \frac{\mathcal{T}_\mathrm{ic}}{3 f(r_\mathrm{ic})}  a \mathcal{P}^{-1} Ra^b \right]^{-1}	\label{PhiPlumes}
\end{equation}
in the high-$\mathcal{P}$ regime.

\subsection{Today's inner core regime and rate of melt production}

The inner core dynamic regime depends mostly on the value of its non-dimensional age $\mathcal{T}_\mathrm{ic}$ and of $\mathcal{P}$, both parameters being very poorly constrained.
The value of $\mathcal{T}_\mathrm{ic}$ dictates whether the inner core has a stable or unstable temperature profile, and the parameter $\mathcal{P}$ determines the convection regime if the inner core is unstable against thermal convection.
Other parameters have a comparatively small influence on the inner core dynamics, and on the value of $\Phi_B$.
With this idea in mind, it is useful to rewrite the Rayleigh number as a function of $\mathcal{P}$ and $\mathcal{T}_\mathrm{ic}$ :
\begin{align}
Ra &= \frac{\alpha \rho _s g_\mathrm{icb} S r_\mathrm{ic}^5}{6 \kappa ^2 \eta } \nonumber  \\
       &= \mathcal{A} \left[ f(r_\mathrm{ic})\,\mathcal{T}_\mathrm{ic}^{-1} - 1  \right]  \mathcal{P},\quad \text{with}\ \mathcal{A}= \frac{\alpha^2 \rho _s^2\, g_\mathrm{icb}  r_\mathrm{ic}^3 T}{2\, k\, \Delta \rho   \, \tau _\phi }. \label{Ra_P_T}
\end{align}
The exact value of the pre-factor $\cal{A}$ affects the value of $\Phi_B$, but not the inner core regime (stably stratified inner core, translation, or  plume convection) which is determined by $\mathcal{P}$ and $\mathcal{T}_\mathrm{ic}$.
The uncertainty on $\mathcal{A}$ comes mostly from the uncertainty on $\tau_\phi$, which is difficult to estimate without a better understanding of the dynamics of the F-layer.
If $\mathcal{P}$ and $\mathcal{T}_\mathrm{ic}$ are kept constants,  changing $\mathcal{A}$ by an order of magnitude would change the translation velocity and melting rates (in both regimes) by a factor of $\sim 3$.
Figure \ref{MeltingRateRegime} shows $\Phi_B$  as a function of $\mathcal{T}_\mathrm{ic}$ and $\mathcal{P}$, calculated from Equation \eqref{PhiTranslation} and Equation \eqref{PhiPlumes} with $Ra$ given by Equation \eqref{Ra_P_T} and $\mathcal{A}=3\ 10^5$ (corresponding to parameters values given in table \ref{parameters}). 
Figure \ref{MeltingRateRegime} serves both as a regime diagram for the inner core, and as a predictive map for $\Phi_B$ and the likelihood of the development of a stratified layer at the base of the outer core.

The inner core has currently an unstable thermal profile only if $\mathcal{T}_\mathrm{ic}$ is smaller than $\simeq 0.87$.
The mode of thermal convection then depends on $\mathcal{P}$, with the translation regime (small $\mathcal{P}$) being the most efficient at producing melt.
Plume convection generates less melt, but the rate of melt production still remains significant as long as $\mathcal{P}$ is not too large ($\eta$ not too small).
The critical value of $\Phi_B=0.8$ (white contour in Figure \ref{MeltingRateRegime}) suggested by the experiments of \cite{Alboussiere2010} is almost always reached in the translation regime, but only in a small part of the parameter space in the plume convection regime.

\section{Compositional effects}
\label{CompositionalFeedback}

We have so far left aside the possible effects of the compositional evolution of the outer and inner core on the inner core dynamics. 
We will argue here that the development of an iron rich layer above the inner core can have a possibly important positive feedback on inner core convection:
irrespectively of the exact mechanism at the origin of the F-layer \citep{Gubbins08,Alboussiere2010,Gubbins2011}, its interpretation as an iron rich layer implies a  decrease with time of light elements concentration in the liquid just above the ICB.
This in turn implies that the newly crystallized solid is increasingly depleted in light elements, and intrinsically denser, which may drive compositional convection in the inner core.
The reciprocal coupling between the inner core and the F-layer may create a positive feedback loop which can make the system (inner core + F-layer) unstable. The mechanism releases more gravitational energy than purely radial inner core growth with no melting, and should therefore be energetically favored.

We note $c^s$ and $c^l$ the light element concentration in the inner and outer core, respectively,  $c_\mathrm{icb}^{s,l}$ their values at the ICB, and $\dot c_\mathrm{icb}^{s,l} = d c_\mathrm{icb}^{s,l}/dt$ their time derivatives at the ICB.
The concentration in the liquid and solid sides of the ICB are linked by the  partition coefficient $k$, $c_\mathrm{icb}^s = k\,c_\mathrm{icb}^l$.
Introducing $\tilde c = c - c_\mathrm{icb}^s$, the equation of transport of light element can be written as
\begin{equation}
\frac{D \tilde c}{Dt} = \kappa_c \nabla^2 \tilde c + S_c, \quad \tilde c(r_\mathrm{ic})=0,
\end{equation}
with
\begin{equation}
S_c = - \dot c_\mathrm{icb}^s = - k\, \dot c_\mathrm{icb}^l - c_\mathrm{icb}^l \frac{d k}{dt},
\end{equation}
which is an exact analogue of the potential temperature transport equation \eqref{entropy_ALA2}. 
The only - but important - difference is that the source term $S_c$ is a dynamical quantity which depends on the convective state of the inner core and on the dynamics of the F-layer rather than being externally imposed like the effective heating rate $S$, which means that the dynamics of the inner core and F-layer must be considered simultaneously.

In general, the fact that the thermal and compositional diffusivities are different can be of importance, and would lead to double-diffusive type convection. 
But this is not the case in the translation regime, for which diffusion does not play any role as long as the translation rate is large enough (\textit{i.e.} if the P\'eclet number $Pe=V r_\mathrm{ic}/\kappa \gg 1$).
Thanks to the potential temperature/composition analogy noted above, the translation model developed for thermal convection can  be extended to include compositional effects, the translation rate being given by
 \begin{equation}
V = \left[ \frac{1}{5}\frac{\rho_s}{\Delta \rho} \frac{(\alpha S + \alpha_c\, S_c)}{\tau_\phi} \right]^{1/2} r_\mathrm{ic}  \label{TranslationComposition}
\end{equation}
when compositional effects are accounted for.
We therefore need to compare the  magnitudes of $\alpha S$ and $\alpha_c\,S_c=-\alpha_c\, k\, \dot c_\mathrm{icb}^l-\alpha_c\,c_\mathrm{icb}^l\, dk/dt$.  
Assuming \citep{Gubbins08} that the light element concentration at the base of the F-layer is currently about twice smaller than the outer core mean concentration, $c_\mathrm{oc}\simeq 5$ wt.\%, we obtain $\dot c_\mathrm{icb}^l\sim -0.5\, c_\mathrm{oc}/\tau_\mathrm{ic}\sim -10^{-18}$ wt.\%.s$^{-1}$ with $\tau_\mathrm{ic}\sim 1$ Gy. 
With $\alpha_c \simeq 1$, this gives $-\alpha_c\, k\, \dot c_\mathrm{icb}^l \sim 10^{-19}$ s$^{-1}$ if $k\simeq 0.1$ and $-\alpha_c\, k\, \dot c_\mathrm{icb}^l \sim 10^{-20}$ s$^{-1}$ if $k\simeq 0.01$, 
which is similar or larger than the thermal contribution  $\alpha\, S \sim 10^{-5}\times 10^{-15}\sim 10^{-20}$~s$^{-1}$.
The term $-\alpha_c\,c_\mathrm{icb}^l\, dk/dt$ might be positive as well. 
According to calculations by \cite{gubbins2013}, the variation with temperature of the partitioning behavior of Oxygen can  produce an unstable compositional gradient.
As discussed in \cite{Alboussiere2010} and \cite{Deguen2011a}, the effective partition coefficient may also decrease with time because of dynamical reasons (the efficiency of melt expulsion from the inner core increases with inner core size), which would also imply that this term is positive.

There  is an additional feedback, this time negative, which comes from the effect of composition on the solidification temperature, which increases with decreasing light element concentration. 
The decreasing light element concentration at  the base of the F-layer implies that the ICB temperature decreases with time at a slower rate than if the composition is fixed, which results in a smaller effective heating rate $S$ [Equation \eqref{DefS}]. 
For a fixed inner core growth rate, this decreases the ICB cooling rate by an amount equal to $-m_c \, \dot c_\mathrm{icb}^l$, where $m_c=\partial T_s/\partial c\sim -10^4$ K \citep{Alfe2002} is the liquidus slope at the inner core boundary pressure and composition.
This adds a term $-\alpha\, m_c \, \dot c_\mathrm{icb}^l$ in Equation \eqref{TranslationComposition}.
If only one light element is considered, the ratio of the stabilizing term $\alpha\, m_c \, \dot c_\mathrm{icb}^l$ over the destabilizing term $-\alpha_c\, k\, \dot c_\mathrm{icb}^l$ is $\sim \alpha\, m_c/(\alpha_c\, k)\sim -0.1/k$.
The two terms are of the same order of magnitude if $k\sim 0.1$, but the negative feedback dominates if $k$ is smaller.

The above estimates are clearly uncertain, and a  dynamical model of the F-layer will be required for 
assessing in a self-consistent way the effect of the development of the F-layer on inner core convection.
There are several feedbacks of the formation of an F-layer on inner core convection, either positive or negative, and it is not clear yet whether the net effect would be stabilizing or destabilizing. 
Still, it does suggest that the effect could be important, and worth considering in more details.

\section{Summary and conclusions}
\label{SectionSummary}

\begin{table*}
\caption{Summary of theoretical results and scaling laws for the translation ($\mathcal{P} \lesssim 29$) and plume convection ($\mathcal{P} \gg 29$) regimes.
In the plume convection regime, the value of $\beta$ obtained by fitting the numerical outputs to our scaling theory is $\beta = -0.238\pm0.003$.
}
\begin{center}
\begin{tabular*}{0.7\linewidth}{@{}@{\extracolsep{\fill}}p{5cm}ll}
\toprule
						&	Translation regime 	&	Plume convection regime \\
						&	$\mathcal{P} \lesssim 29$  	&	$\mathcal{P} \gg 29$	\\
\midrule
Onset					&    $\left(\dfrac{Ra}{\mathcal{P}}\right)_{\!\!c} = \dfrac{175}{2}$	&	$Ra_c = 1545.6$	\\
\midrule
Velocity scaling	, $V$ or $U_\mathrm{rms}$			&  $\dfrac{\kappa}{r_\mathrm{ic}} \sqrt{\dfrac{6}{5}\dfrac{Ra}{\mathcal{P}}}$	& $0.96 \dfrac{\kappa}{r_\mathrm{ic}} Ra^{2+7\beta}$	\\
Rate of melt production, $\dot M$	&  $\dfrac{1}{4}\dfrac{\kappa}{r_\mathrm{ic}} \sqrt{\dfrac{6}{5}\dfrac{Ra}{\mathcal{P}}}$	&	$0.46\dfrac{\kappa}{r_\mathrm{ic}} Ra^{1+2\beta} \mathcal{P}^{-1}$\\
$\left< \Theta \right>$				&	$\dfrac{S r_\mathrm{ic}^2}{\kappa} \left(\dfrac{10}{3}\dfrac{\mathcal{P}}{Ra}\right)^{1/2}$			&$2.9 \dfrac{S r_\mathrm{ic}^2}{\kappa} Ra^\beta$	\\
Number of plumes per unit area, $N$	& -	& $\dfrac{0.07}{r_\mathrm{ic}^2} Ra^{-2-10\beta}$\\
Strain rate $\dot \epsilon\sim \dfrac{U_\mathrm{rms}}{\lambda} \sim \sqrt{N} U_\mathrm{rms}$	& - & $0.25 \dfrac{\kappa}{r_\mathrm{ic}^2} Ra^{1+2\beta}$  \\
\bottomrule
\end{tabular*}
\end{center}
$Ra = \dfrac{\alpha \rho _s g_\mathrm{icb} S r_\mathrm{ic}^5 }{6 \kappa ^2 \eta }$, ${\cal{P}} = \dfrac{\Delta \rho \, g_\mathrm{icb} \, r_\mathrm{ic} \, \tau _\phi }{\eta }$.
\label{ScalingSummary}
\end{table*}

Inner core translation can potentially  explain a significant part of the inner core structure, but its existence depends critically on the value of a number of poorly constrained parameters. 
In this paper, we have studied in details the conditions for and dynamics of inner core thermal convection when melting and solidification at the ICB are allowed. 
We summarize here the main results and implications of our work~: 

\begin{enumerate}
\item If the inner core is  convectively unstable, linear stability analysis (section \ref{linear_stability}), asymptotic calculations (section \ref{Small_P}), and direct numerical simulations (section \ref{Numerical_results}) consistently show that the convection regime depends mostly on a non-dimensional number, the "phase change number" $\mathcal{P}$, characterizing the resistance to phase change [Equation \eqref{P}].
The convective translation mode dominates only if $\mathcal{P}<29$, which requires that  the inner core viscosity is larger than a critical  value estimated to be $\sim 3 \times 10^{18}$ Pa~s.
If $\mathcal{P}$ is larger (smaller viscosity), melting and solidification at the ICB have only a small dynamical effect, and convection takes the usual form of low Prandtl number internally heated convection, with a one-cell axisymmetric mode at the onset, and chaotic plume convection if the Rayleigh number is large.
\item With published estimates of the inner core viscosity ranging from $10^{11}$ Pa~s to $10^{22}$ Pa~s \citep{Yoshida1996,Buffett1997,VanOrman04,Mound2006,Koot2011,Reaman2011,Reaman2012}, the question of which mode would be preferred is open
(although we note that the latest estimate from mineral physics, $10^{20}-10^{22}$ Pa~s \citep{Reaman2012}, would put the inner core, if unstably stratified, well within the translation regime).

\item
The two convection regimes have been characterized in details in sections \ref{linear_stability} to \ref{application};
a summary of theoretical results and scaling laws for useful dynamical quantities (RMS velocity, rate of melt production, mean potential temperature, number of plumes per unit area, and strain rate) is given in table \ref{ScalingSummary}.
If the inner core is unstably stratified, the rate of melt production predicted by our models is always large enough to produce an iron-rich layer at the base of the outer, according to \cite{Alboussiere2010}'s experiments, if the inner core is in the translation regime (Figure \ref{MeltingRateRegime}).
In the plume convection regime, the rate of melt production can still be significant if $\mathcal{P}$ is not too large ($\eta$ not too small).

\item Being driven by buoyancy, a prerequisite for the existence of convective translation is that an unstable density profile is maintained within the inner core.
Thermal convection requires  that a superadiabatic temperature profile is maintained with the inner core, which is highly dependent on the core thermal history and inner core thermal conductivity.
With $k=36$ W.m$^{-1}$.K$^{-1}$ as proposed by \cite{Stacey08}, this would be very likely \citep{Buffett2009,Deguen2011a}.
However, several independent groups \citep{sha2011,deKoker2012,pozzo2012} have recently argued for a much higher core thermal conductivity, around 150 W.m$^{-1}$.K$^{-1}$ or higher. 
This would make thermal convection in the inner core, whether in the translation mode or in the plume convection mode,  impossible unless the inner core is very young ($\simeq 300$ My or less, which would require a probably excessively high CMB heat flux).
\item Compositional convection might be a viable alternative to thermal convection, either because the temperature dependency of the light elements partitioning behavior  can produce an unstable compositional profile  \citep{gubbins2013}, or because of a possibly positive feedback of the development of the F-layer on inner core convection.
As proposed in section \ref{CompositionalFeedback}, the formation of an iron-rich layer at the base of the outer core over the history of the inner core implies that the inner core crystallizes from a source which is increasingly  depleted in light elements.
This in turn implies that the newly crystallized solid is increasingly depleted in light element, which results in an unstable density profile.
Whether this positive feedback is strong enough to overcome the stabilizing effect of a possibly subadiabatic temperature profile depends on the dynamics of the F-layer, and further work is needed to test this idea.
\end{enumerate}

\begin{acknowledgments}

We would like to thank the two anonymous referees for many helpful comments and suggestions.
All the computations presented in this paper were performed at the Service Commun de Calcul Intensif de l'Observatoire de Grenoble (SCCI).
R.~D. was  supported by grant EAR-0909622 and Frontiers in Earth System Dynamics grant EAR-1135382 from the National Science Foundation.
T.~A. was supported by the ANR (Agence National de la Recherche) within the CrysCore project ANR-08-BLAN-0234-01, and by the program PNP of INSU/CNRS.
P.~C. was supported by the program PNP of INSU/CNRS.

\end{acknowledgments}

\appendix

\section{Linear stability analysis  }
\label{AppendixLinearStability}

We investigate here the linear stability of the system of equations  describing thermal convection in the inner core with phase change at the ICB, as derived in section \ref{gov_eqns}. 
The calculation given here is a generalization of the linear stability analysis of thermal convection in an internally heated sphere given by \cite{Chandrasekhar1961}.

We assume constant $Ra$ and $\mathcal{P}$.
The basic state of the problem is then
\begin{align}
\bar \Theta &= 1 - r^2, \label{AppendixTbase} \\  
\bar{ \mathbf{u}} &= \mathbf{0},
\end{align}
which is the steady conductive solution of the system of equation developed in section \ref{gov_eqns}.
We investigate the stability of this conductive state against infinitesimal perturbations of the temperature and velocity fields.
The temperature field is written as the sum of the conductive temperature profile given by equation \eqref{AppendixTbase} and infinitesimal disturbances $\tilde \Theta$, $\Theta(r,\theta,\phi,t) = \bar \Theta(r) + \tilde \Theta(r,\theta,\phi,t)$. 
The velocity field perturbation is noted $\tilde{\mathbf{u}}(r,\theta,\phi,t)$, and has an associated poloidal scalar $\tilde P(r,\theta,\phi,t)$.
We expand the temperature and poloidal disturbances in spherical harmonics,
\begin{align}
\tilde \Theta &= \sum_{l=0}^\infty \sum_{m=-l}^l \tilde t_l^m(r) Y_l^m(\theta,\phi)\, \mathrm{e}^{\sigma_l t},   \label{AppendixTDecomposition}\\
\tilde P &=\sum_{l=1}^\infty \sum_{m=-l}^l \tilde p_l^m(r) Y_l^m(\theta,\phi)\, \mathrm{e}^{\sigma_l t}, \label{AppendixPDecomposition}
\end{align}
where $\sigma_l$ is the growth rate of the degree $l$ perturbations.

The only non-linear term in the system of equations is the advection of heat $\mathbf{u}\cdot \nabla T$, which is linearized as 
\begin{equation}
\tilde u_r \frac{\partial \bar \Theta}{\partial r} = -2 r \tilde u_r = - 2 L^2 \tilde P .  
\end{equation}
The resulting linearized transport equation for the potential temperature disturbance is
\begin{equation} 
\frac{\partial \tilde \Theta}{\partial  t} = {\nabla}^2 \tilde \Theta + 2 L^2 \tilde P + 6.
\end{equation} 
Using the decompositions \eqref{AppendixTDecomposition} and \eqref{AppendixPDecomposition}, the linearized system of equations is then, for $l\geq 1$,
\begin{align}
Ra\, \tilde t_l^m &= \mathcal{D}_l^2 \tilde p_l^m,\\
\left( \sigma_l -  \mathcal{D}_l \right) \tilde t_l^m &= 2 l (l+1) \tilde p_l^m   \label{HeatMarg}
\end{align}
with the boundary conditions given by Equations \eqref{stressfreePlm} and \eqref{normalstressPlm}, with $\tilde t_l^m(r=1)=0$.

We expand the temperature perturbations $\tilde t_l^m(r)$ as a series of spherical Bessel functions of the first kind $j_l$,
\begin{equation}
\tilde t_l^m = \sum_i A_{l,i} j_l(\alpha_{l,i} r)   \label{Def_t_l_m}
\end{equation}
The spherical Bessel functions are defined as
\begin{equation}
j_l(r) = \sqrt{\frac{\pi}{2 r}} J_{l+\frac{1}{2}}(r),
\end{equation}
where $J$ denotes Bessel functions of the first kind. $\alpha_{l,i}$ is the $i$th zero of $J_{l+\frac{1}{2}}$, and therefore of $j_l$ as well.
The functions $j_l(\alpha_{l,i} r)$ for $i=1,2,...,\infty$ and a given $l$ form a complete set of orthogonal functions on $[0,1]$, and satisfy the orthogonality relation
\begin{equation}
\int_0^1 r^2 j_l(\alpha_{l,i}r) j_l(\alpha_{l,j}r) dr = \frac{\delta_{i,j}}{2} \left[  j_{l+1}(\alpha_{l,j}) \right]^2. \label{OrthogonalityRelation}
\end{equation}
The spherical Bessel functions are eigenfunctions of the operator $\mathcal{D}_l$, such that
\begin{equation}
\mathcal{D}_l j_l(\alpha_{l,i} r) = - \alpha_{l,i}^2\, j_l(\alpha_{l,i} r).
\end{equation}

Writing the poloidal scalar perturbations $\tilde p_l^m$ as
\begin{equation}
\tilde p_l^m = \sum_i A_{l,i} p_{l,i},  \label{Def_p}
\end{equation}
the functions $p_{l,i}$ are solutions of 
\begin{equation}
Ra\,  j_l(\alpha_{l,i} r) = \mathcal{D}_l^2 p_{l,i},
\end{equation}
which has a general solution of the form
\begin{equation}
p_{l,i} = \frac{Ra}{\alpha_{l,i}^4}  j_l(\alpha_{l,i} r) + B_{l,i} r^l + C_{l,i} r^{l+2}.   \label{GeneralSolution_p} 
\end{equation}

We now use the boundary conditions at $r=1$ to find the constants $B_{l,i}$ and $C_{l,i}$.
The  condition of zero tangential stress  [Equation \eqref{stressfreePlm}] can be rewritten as
\begin{align}
\mathcal{D}_l \tilde p_l^m - 2 \frac{d \tilde p_l^m}{dr} + 2 \left[ l(l+1) - 1 \right] \tilde p_l^m= 0, 
\end{align}
which, recalling that $j_l(\alpha_{l,i})=0$ and noting that
\begin{equation}
\mathcal{D}_l \tilde p_l^m = \sum_i A_{l,i} \left[ - \frac{Ra}{\alpha_{l,i}^2} j_l(\alpha_{l,i} r) + C_{l,i} (4l+6) r^l  \right],
\end{equation}
gives
\begin{equation}
C_{l,i} = \frac{1-l^2}{l(l+2)} B_{l,i} + \frac{1}{l(l+2)} \frac{Ra}{\alpha_{l,i}^3}  j_l'(\alpha_{l,i}). \label{C_{l,i}}
\end{equation}
From the continuity of the normal stress at $r=1$ [Equation \eqref{normalstressPlm}], we obtain
\begin{equation}
\begin{split}
&   B_{l,i} = - \left[ 2(l-1) + \mathcal{P} \right]^{-1} \left[\frac{1}{l(l+1)}  +  \frac{2}{\alpha_{l,i}^2}\right]  \frac{j_l'(\alpha_{l,i})}{\alpha_{l,i}} Ra \\
& - C_{l,i} + \left[ 2(l-1) + \mathcal{P} \right]^{-1} \frac{6}{l} C_{l,i}  .  \label{BC2}
\end{split}
\end{equation}
The derivative of $j_l$ which appears in equations \eqref{C_{l,i}} and \eqref{BC2} can be evaluated from the recurrence relation 
\begin{equation}
\frac{n}{r} j_n - \frac{d j_n}{dr} = j_{n+1} \label{Rec1}
\end{equation}
\citep{abramovich1965}. Recalling that $j_l(\alpha_{l,i})=0$, equation \ref{Rec1} with $n=l$ gives
\begin{equation}
 j_l'(\alpha_{l,i}) = - j_{l+1}(\alpha_{l,i}).
\end{equation}
Inserting Equation \eqref{C_{l,i}} in Equation \eqref{BC2}, we obtain
\begin{equation}
\begin{split}
&    \left\{ 4 l (l+1) - 2 + (2l+1) \mathcal{P} -  \frac{6}{l} \right\}B_{l,i} = \\
&\left\{\frac{l+2}{l+1}  +   \left[2 (l^2+3l-1) + \mathcal{P} - \frac{6}{l} \right] \frac{1 }{\alpha_{l,i}^2}  \right\}  \frac{j_{l+1}(\alpha_{l,i})}{\alpha_{l,i}} Ra,  
\end{split}
\end{equation}
which we rewrite as
\begin{equation}
B_{l,i} =  \left( q_1^l(\mathcal{P}) +  \frac{q_2^l(\mathcal{P})}{\alpha_{l,i}^2} \right)  \frac{j_{l+1}(\alpha_{l,i})}{\alpha_{l,i}} Ra,
\end{equation}
where
\begin{align}
q_1^l(\mathcal{P}) &= \frac{l+2}{l+1} \left[ 4 l (l+1) - 2 + (2l+1) \mathcal{P} -  \frac{6}{l} \right]^{-1}, \label{q1} \\
q_2^l(\mathcal{P}) &=  \frac{2 (l^2+3l-1) + \mathcal{P} - \frac{6}{l} }{ 4 l (l+1) - 2 + (2l+1) \mathcal{P} -  \frac{6}{l} }. \label{q2}
\end{align}
With this expression for $B_{l,i}$, the constants $C_{l,i}$ are given by
\begin{equation}
C_{l,i} =  -\left(  q_3^l(\mathcal{P}) +   \frac{q_4^l(\mathcal{P})}{\alpha_{l,i}^2} \right)  \frac{j_{l+1}(\alpha_{l,i})}{\alpha_{l,i}} Ra,
\end{equation}
where
\begin{align}
q_3^l(\mathcal{P}) &= \frac{(l^2-1)}{l(l+2)} q_1^l(\mathcal{P}), \label{q3}\\
q_4^l(\mathcal{P}) &= \frac{ (l^2-1) q_2^l(\mathcal{P}) + 1 }{l(l+2)}. \label{q4}
\end{align}

Now, using Equations \eqref{Def_p} and  \eqref{GeneralSolution_p} for $\tilde p_l^m$ and Equation \eqref{Def_t_l_m} for $\tilde t_l^m$,
 the heat equation \eqref{HeatMarg} gives
\begin{equation}
\begin{split}
\sum_i A_{l,i}  \left( \frac{\sigma_l + \alpha_{l,i}^2}{2l(l+1)}  - \frac{Ra}{\alpha_{l,i}^4} \right)& j_l(\alpha_{l,i} r) \\ 
=\sum_i A_{l,i}& \left(  B_{l,i} r^l +  C_{l,i} r^{l+2}  \right). \label{Disp1}
\end{split}
\end{equation}
Multiplying equation \eqref{Disp1} by $r^2 j_l(\alpha_{l,j}r)$ where $j$ is an integer in $[0; \infty[$,  integrating in $r$ over $[0; 1]$, and using the orthogonality relation \eqref{OrthogonalityRelation}, we obtain
\begin{equation}
\begin{split}
A_{l,j} \left( \frac{\sigma_l + \alpha_{l,j}^2}{2l(l+1)}  - \frac{Ra}{\alpha_{l,j}^4} \right)& \frac{1}{2} \left[  j_{l+1}(\alpha_{l,j}) \right]^2 \\
= \sum_i A_{l,i}  B_{l,i} &\int_0^1 r^{l+2} j_l(\alpha_{l,j}r) dr \\
+ \sum_i A_{l,i}   C_{l,i} &\int_0^1 r^{l+4} j_l(\alpha_{l,j}r) dr, \quad (j=1,2,\dots) . \label{LinearSystem1}
\end{split}
\end{equation}
Equation \eqref{LinearSystem1} forms a set of linear homogeneous equations for the constants $A_{l,j}$, which admits non-trivial solutions only if its secular determinant is equal to zero.

Before calculating the secular determinant of the system of equations, we evaluate the two integrals on the right hand side, starting with the integral of $r^{l+2} j_l(\alpha_{l,j}r)$. Using the formula
\begin{equation}
\left( \frac{1}{x}\frac{d}{dx} \right)^m \left[ x^{n+1} j_n(x)  \right] = x^{n-m+1}j_{n-m}(x)
\end{equation}
\citep{abramovich1965} with $m=1$ and $n=k+1$ gives
\begin{equation}
\frac{d}{dx}  \left[ x^{k+2} j_{k+1}(x)  \right] = x^{k+2} j_k(z), \label{DerivativeFormula}
\end{equation}
which allows to write, with $k=l$,
\begin{equation}
\begin{split}
\int_0^1 r^{l+2} j_l(\alpha_{l,j}r) dr &= \frac{1}{\alpha_{l,j}^{l+3}} \int_0^{\alpha_{l,j}} x^{l+2} j_l(x) dx \\
						  &= \frac{1}{\alpha_{l,j}^{l+3}} \left[  x^{l+2} j_{l+1}(x)  \right]_0^{\alpha_{l,j}}= \frac{j_{l+1}(\alpha_{l,j})}{\alpha_{l,j}} .\label{Int_l+2}
\end{split}
\end{equation}
Now, using the recurrence relation
\begin{equation}
j_{n-1}+ j_{n+1} = \frac{2n+1}{r} j_n  \label{Rec2}
\end{equation}
\citep{abramovich1965} with $n=l+1$, we rewrite the integral of $r^{l+4} j_l(\alpha_{l,j}r)$ as
\begin{align}
\int_0^1 r^{l+4} j_l(\alpha_{l,j}r) dr &= \frac{1}{\alpha_{l,j}^{l+5}} \int_0^{\alpha_{l,j}} x^{l+4} j_{l}(x) dx \\
\begin{split}=
       &- \frac{1}{\alpha_{l,j}^{l+5}}  \int_0^{\alpha_{l,j}} x^{l+4} j_{l+2}(x) dx\\
      &+ \frac{2l+3}{\alpha_{l,j}^{l+5}} \int_0^{\alpha_{l,j}} x^{l+3} j_{l+1}(x) dx
      \end{split}
\end{align}
The two integrals on the RHS can be calculated using the relation \eqref{DerivativeFormula} with $k=l+2$ and $k=l+1$, respectively. With further use of the recurrence relation \eqref{Rec2}, we finally obtain
\begin{equation}
\int_0^1 r^{l+4} j_l(\alpha_{l,j}r) dr = \left[1-\frac{4l+6}{\alpha_{l,j}^2}\right] \frac{j_{l+1}(\alpha_{l,j})}{\alpha_{l,j}}.
\end{equation}

With the integrals estimated  above, the system of equations \eqref{LinearSystem1} can be rewritten as
\begin{equation}
\begin{split}
A_{l,j} &\left( \frac{\sigma_l + \alpha_{l,j}^2}{2l(l+1) Ra}  - \frac{1}{\alpha_{l,j}^4} \right) \frac{1}{2} \left[  j_{l+1}(\alpha_{l,j}) \right]^2 \\
=& \sum_i A_{l,i} \left\{\left(  q_3^l(\mathcal{P}) +   \frac{q_4^l(\mathcal{P})}{\alpha_{l,i}^2} \right) \left[\frac{4l+6}{\alpha_{l,j}^2}-1\right]  \right.\\
+& \left.    q_1^l(\mathcal{P}) +  \frac{q_2^l(\mathcal{P})}{\alpha_{l,i}^2} \right\} \frac{j_{l+1}(\alpha_{l,i})}{\alpha_{l,i}}    \frac{j_{l+1}(\alpha_{l,j})}{\alpha_{l,j}},\quad (j=1,2,\dots).
\end{split}
\end{equation}
Introducing $\mathcal{A}_i =( j_{l+1}(\alpha_{l,i})/\alpha_{l,i}^3 ) A_{l,i}$, and dividing by $j_{l+1}(\alpha_{l,j})/\alpha_{l,j}$, we finally obtain
\begin{equation}
\begin{split}
\sum_i \mathcal{A}_i  & \left\{ \left[  q_3^l(\mathcal{P}) \alpha_{l,i}^2 +   q_4^l(\mathcal{P}) \right] \left[1-\frac{4l+6}{\alpha_{l,j}^2} \right]  \right. \\
-&\left[ q_1^l(\mathcal{P}) \alpha_{l,i}^2 +  q_2^l(\mathcal{P}) \right]   \\
+& \left. \left( \frac{\sigma_l \alpha_{l,i}^4 + \alpha_{l,i}^6}{2l(l+1) Ra}  - 1\right) \frac{1}{2}   \delta_{ij}  \right\}=0,\quad (j=1,2,\dots).  \label{SystemEquationsLS}
\end{split}
\end{equation}

This forms an infinite set of linear equations, which admits a non trivial solution only if its determinant is zero :
\begin{equation}
\begin{split}
&\left|\left| \left[  q_3^l(\mathcal{P}) \alpha_{l,i}^2 +   q_4^l(\mathcal{P}) \right] \left[1-\frac{4l+6}{\alpha_{l,j}^2} \right] -\left[ q_1^l(\mathcal{P}) \alpha_{l,i}^2 +  q_2^l(\mathcal{P}) \right]  \right.\right. \\
+& \left.\left. \left( \frac{\sigma_l \alpha_{l,i}^4 + \alpha_{l,i}^6}{2l(l+1) Ra}  - 1\right) \frac{1}{2}   \delta_{ij} \right|\right| =0,  
\end{split}\label{Determinant_l}
\end{equation}
with $i,j=1,2,...$.
Solving equation \eqref{Determinant_l} for a given value of $l$ and $\sigma_l=0$ gives the critical value $Ra_c$ of the Rayleigh number for instability of the $l$ mode as a function of $\mathcal{P}$.
When solving numerically equations \eqref{Determinant_l}, the precision on $Ra_c$ depends on the maximum value of $i$ and $j$ retained in the calculation, but the value of $Ra_c$ converges relatively fast with $i,j$. 

The pattern of the first unstable mode can be calculated by solving the system \eqref{SystemEquationsLS} in $\mathcal{A}_i$ for given $\mathcal{P}$ and $Ra$, which gives $A_{l,i}$ and allows to calculate the poloidal scalar $\tilde{p}_l^m$ from equations \eqref{Def_p} and \eqref{GeneralSolution_p}.	
With $l=1$, we have $q_1^1=1/(2\mathcal{P})$, $q_2^1=1/3$, $q_3^1=0$ and $q_4^1=1/3$, so that the functions $p_{1,i}$ can be written as
\begin{equation}
p_{1,i} =  \frac{ j_1(\alpha_{1,i} r)}{\alpha_{1,i}}  +  \frac{j_2( \alpha_{1,i})}{3}( r - r^{3}) + \frac{ j_2( \alpha_{1,i})\alpha_{1,i}^2}{2\, \mathcal{P}} r,  \label{p_{l,i}}
\end{equation}
and the general form of the $l=1$, $m\in[-1,0,1]$ components of the poloidal scalar is
 \begin{equation}
 \tilde{p}_1^m = \sum_{i=1}^\infty A_{1,i} \left[  \frac{ j_1(\alpha_{1,i} r)}{\alpha_{1,i}}  +  \frac{j_2( \alpha_{1,i})}{3}( r - r^{3}) + \frac{ j_2( \alpha_{1,i})\alpha_{1,i}^2}{2\, \mathcal{P}} r \right].
 \end{equation}
 To a good approximation, the first unstable mode is given (to within a multiplicative constant) by keeping  only the $i=j=1$ term, 
\begin{equation}
\tilde{P} \simeq \left\{   \frac{ j_1(\alpha_{1,1} r)}{\alpha_{1,1}}  +  \frac{j_2( \alpha_{1,1})}{3}( r - r^{3}) + \frac{ j_2( \alpha_{1,1})\alpha_{1,1}^2}{2\, \mathcal{P}} r  \right\} \cos \theta.
\end{equation}

\section{Translation rate at $\mathcal{O}(\mathcal{P})$ }
\label{Appendix_Tr}

In order to estimate the translation velocity at $\mathcal{O}(\mathcal{P})$, we need to determine the parameter $A$ in the $\mathcal{O}(\mathcal{P})$ expansion of $p_1$ (equation \eqref{p10gene}), which was left undetermined.
To do so, we need to consider the thermal field at $\mathcal{O}(\mathcal{P})$ and the velocity field at $\mathcal{O}(\mathcal{P}^2)$.
This is more challenging because, owing to the non-linearity of the heat equation, coupling of higher order components of the temperature and velocity fields contribute to the $l=1$ component of the temperature field at $\mathcal{O}(\mathcal{P})$, and to the $l=1$ component of the velocity field at $\mathcal{O}(\mathcal{P}^2)$.

As before, we consider a steady state approximation of the heat equation where advection and internal heating balance,
\begin{equation}
\mathbf{u}\cdot \nabla \Theta = u_r \frac{\partial \Theta}{\partial r} + \frac{u_\theta}{r} \frac{\partial \Theta}{\partial \theta} = 6.   \label{HeatEquation_ApB}
\end{equation}
Using Legendre polynomial expansions of the poloidal and temperature field, 
\begin{equation}
\Theta = \sum_{l=0}^\infty t_l P_l(\cos \theta), \quad P'  = \sum_{l=0}^\infty p_l P_l(\cos \theta),
\end{equation}
equation \ref{HeatEquation_ApB} can be rewritten as
\begin{equation}
\begin{split}
6=&\left( \sum_l l(l+1) \frac{p_l}{r} P_l (\cos\theta) \right)\times\left( \sum_l \frac{d t_l}{d r} P_l (\cos\theta) \right)\\
+ \frac{1}{r} &\left( \sum_l \frac{1}{r} \frac{d}{dr} \left( r p_l  \right) \frac{d P_l (\cos\theta)}{d \theta} \right)\times\left( \sum_l t_l \frac{d P_l (\cos\theta)}{d \theta} \right) \label{HeatEq_spectral}
\end{split}
\end{equation}
(We can use Legendre polynomials rather than full spherical harmonics  because we restrict the calculation to axisymmetric flows. This gives slightly simpler expressions.)
The $l=1$ and $l=2$ component of the temperature field being much larger than higher order components (with odd $l$ components being zero), we consider only the $l=1$ and $l=2$ terms.
Multiplying  equation \eqref{HeatEq_spectral} by $\sin \theta$ and integrating over $[0\ \pi]$ in $\theta$ then gives
\begin{equation}
\begin{split}
12 = & \frac{4}{3} \frac{p_1}{r} \frac{dt_1}{dr} + \frac{12}{5} \frac{p_2}{r} \frac{dt_2}{dr} \\
      +  & \frac{4}{3} \frac{1}{r^2} \frac{d}{dr} \left( r p_1 \right) t_1 + \frac{12}{5}  \frac{1}{r^2} \frac{d}{dr} \left( r p_2 \right) t_2
\end{split}
\end{equation}
which can be rewritten as
\begin{equation}
3 r^2=   \frac{d}{dr}\left( \frac{1}{3} r p_1   t_1 + \frac{3}{5} r p_2   t_2 \right) .  \label{HeatEq_12_diff}
\end{equation}
Integrating equation \eqref{HeatEq_12_diff} gives
\begin{equation}
r^3 + \mathrm{cst} = \frac{1}{3} r p_1   t_1 + \frac{3}{5} r p_2   t_2 .  \label{HeatEq_12}
\end{equation}

We now expand the Legendre components of the temperature and poloidal scalar fields as
\begin{equation}
\begin{split}
t_1 &= \frac{6}{V_0} \left[ r + \hat{t}_{1,1} \mathcal{P} + \mathcal{O}(\mathcal{P}^2)\right], \quad  t_2 &=& \frac{1}{V_0}\left[ \hat{t}_{2,0} + \mathcal{O}(\mathcal{P}] \right)\\
p_1 &= \frac{V_0}{2}\left[ r + \hat{p}_{1,1} \mathcal{P} + \mathcal{O}(\mathcal{P}^2) \right], \quad  p_2 &=& V_0 \left[ \hat{p}_{2,1} \mathcal{P} + \mathcal{O}(\mathcal{P}^2] \right)
\end{split}
\end{equation}
and insert these expressions in equation \eqref{HeatEq_12}. The zeroth order terms cancel, and equation \eqref{HeatEq_12} then writes
\begin{equation}
0 =  \left( r \hat{p}_{1,1}  +  r   \hat{t}_{1,1} + \frac{3}{5}  \hat{p}_{2,1}  \hat{t}_{2,0}  \right) \mathcal{P}  + \mathcal{O}(\mathcal{P}^2 )
\end{equation}
which implies that
\begin{equation}
\hat{t}_{1,1} = - \hat{p}_{1,1}  - \frac{1}{r} \frac{3}{5}  \hat{p}_{2,1}  \hat{t}_{2,0}.   \label{t_1_1}
\end{equation}

\subsection{$l=2$ components of the thermal field and velocity field}

We now calculate the $l=2$ component of the temperature field at zeroth order in $\mathcal{P}$, which will then be used to find the $l=2$ component of the velocity field at $\mathcal{O}(\mathcal{P})$.

It will be useful to first note that
\begin{equation}
\mathcal{D}_l^2 (r^a) = \left[a(a+1)-l(l+1)  \right]\left[ (a-2)(a-1) -l(l+1) \right]r^{a-4},
\end{equation}
from which we find that
\begin{equation}
\left( \mathcal{D}_1^2  \right)^{-1} (r^a) = \frac{r^{a+4}}{(a+6)(a+4)(a+3)(a+1)} 
\end{equation}
and
\begin{equation}
\left( \mathcal{D}_2^2  \right)^{-1} (r^a) = \frac{r^{a+4}}{(a+7)(a+5)(a+2)a } .
\end{equation}

The $l=2$ component of the  temperature field at zeroth order in $\mathcal{P}$ can be found by direct integration of the temperature field given by Equation \eqref{TemperatureTranslation} :
\begin{equation}
\begin{split}
t_2 =& \frac{5}{2} \frac{6}{V_0} \int_0^\pi \sqrt{1-r^2 \sin^2 \theta}\, P_2(\cos \theta) \sin \theta d\theta \\
                     =& \frac{5}{2}  \frac{9}{4V_0} \left\{ \frac{1}{r^2}-\frac{1}{3} + \frac{1-r^2}{2r}\left( \frac{1}{3}+\frac{1}{r^2} \right) \log \left( \frac{1-r}{1+r}  \right) \right\}\\
                     =& \frac{15}{8V_0} \sum_{k=1}^{+\infty} \left( \frac{1}{2k-1} + \frac{2}{2k+1} - \frac{3}{2k+3} \right)  r^{2k}\\
                     =& \frac{1}{V_0} \sum_{k=1}^{+\infty} \alpha_k  r^{2k}
\end{split}
\end{equation}
with 
\begin{equation}
\alpha_k = \frac{30 k}{(2k + 3)(2k + 1)(2k - 1)}.
\end{equation}

From this, we can calculate the associated velocity field,
\begin{align}
p_2 &= Ra \left( \mathcal{D}_2^2 \right)^{-1} t_2 \\
	   &=  \frac{Ra}{V_0} \left( \mathcal{D}_2^2 \right)^{-1} \left(  \sum_{k=1}^{+\infty} \alpha_k  r^{2k} \right)\\
	   &= \frac{5}{6} V_0 \mathcal{P} \left( \mathcal{D}_2^2 \right)^{-1}  \left( \sum_{k=1}^{+\infty} \alpha_k  r^{2k} \right) \\
	   &=  V_0 \mathcal{P}   \sum_{k=1}^{+\infty} \frac{5}{6} \frac{  \alpha_k}{(2k+7)(2k+5)(2k+2)2k}  r^{2k+4}
\end{align}
The general solution for $\hat{p}_{2,1}$ is
\begin{equation}
\hat{p}_{2,1} = A_2 r^2 + B_2 r^4 +   \sum_{k=1}^{+\infty} \beta_k r^{2k+4},
\end{equation}
where
\begin{align}
\beta_k &= \frac{5}{6} \frac{  \alpha_k}{(2k+7)(2k+5)(2k+2)2k} \\
	    &=  \frac{ 25 }{2(2k+7)(2k+5)(2k + 3)(2k+2)(2k + 1)(2k - 1)}
\end{align}
The constants $A_2$ and $B_2$ have to be determined from the boundary conditions.
The stress free condition gives
\begin{equation}
3 A_2 + 8 B_2 +  \sum_{k=1}^{+\infty} \beta_k \left[ 2 + (k+2)(2k+3) \right] = 0    \label{BC_p2_1}
\end{equation}
and the continuity of normal stress gives, ignoring $\mathcal{O}(\mathcal{P})$ terms, 
\begin{equation}
-15 A_2 - 21 B_2 +  \sum_{k=1}^{+\infty} \beta_k (2 k + 7)(2k^2 + 2k - 3) = 0. \label{BC_p2_2} 
\end{equation}
From equations \eqref{BC_p2_1} and \eqref{BC_p2_2}, we obtain
\begin{equation}
B_2 = - \frac{1}{19}  \sum_{k=1}^{+\infty}  (k + 1)(4k^2 + 24k + 19)  \beta_k \simeq -0.0211
\end{equation}
and 
\begin{equation}
A_2 =  \frac{1}{57}  \sum_{k=1}^{+\infty} k (32k^2+186 k +211 ) \beta_k \simeq 0.0346 .
\end{equation}

\subsection{$l=1$ temperature field at $\mathcal{O}(\mathcal{P})$ and velocity field at $\mathcal{O}(\mathcal{P}^2)$}

Inserting in equation \eqref{t_1_1} the expression found above for the $l=2$ component of the velocity field, $\hat{t}_{1,1}$ is now given by
\begin{equation}
\hat{t}_{1,1} =  - \hat{p}_{1,1} - \frac{3}{5} \left( A_2 r  + B_2 r^3 +   \sum_{k=1}^{+\infty} \beta_k r^{2k+3} \right)  \left(  \sum_{k=1}^{+\infty} \alpha_k  r^{2k} \right) .
\end{equation}
After some rearrangements, we obtain 
\begin{equation}
\begin{split}
\hat{t}_{1,1} =  - \hat{p}_{1,1} - \frac{3}{5}   \sum_{k=0}^{+\infty} \left(A_2 \alpha_{k+1} + B_2 \alpha_k + \gamma_k - \beta_0 \alpha_k \right) r^{2k+3}, 
\end{split}
\end{equation}
where
\begin{equation}
\gamma_k = \sum_{i=0}^{k} \alpha_i \beta_{k-i}.
\end{equation}

We can now determine the $l=1$ flow field by integrating the Stokes equation with the above temperature field.
Noting
\begin{equation}
p_1 = \frac{V_0}{2}\left[ r + \hat{p}_{1,1} \mathcal{P} +  \hat{p}_{1,2} \mathcal{P}^2 + \mathcal{O}(\mathcal{P}^3) \right] 
\end{equation}
the second order contribution is given by
\begin{equation}
\frac{V_0}{2} \mathcal{P}^2  \hat{p}_{1,2} = \frac{6}{V_0}\mathcal{P} Ra \left( \mathcal{D}_1^2 \right)^{-1} \hat{t}_{1,1},
\end{equation}
or
\begin{equation}
\hat{p}_{1,2} = 10 \left( \mathcal{D}_1^2 \right)^{-1} \hat{t}_{1,1} + D r + E r^3.
\end{equation}
We obtain
\begin{equation}
\begin{split}
\hat{p}_{1,2} &= -\frac{1}{28} A r^5 - \frac{5}{756} B r^7 - \frac{5}{2376} C r^9 \\
&- \frac{30}{5}  \sum_{k=0}^{+\infty} \frac{A_2 \alpha_{k+1} + (B_2 - \beta_0) \alpha_k + \gamma_k }{(2k+9)(2k+6)(2k+7)(2k+4)} r^{2k+7}\\
&+ D r + E r^3
\end{split}
\end{equation}
which gives
\begin{equation}
\begin{split}
\hat{p}_{1,2} &= -\frac{1}{28} A r^5 + \frac{25}{31752} r^7 - \frac{5}{66528}  r^9 \\
&- \frac{30}{5}  \sum_{k=0}^{+\infty} \frac{A_2 \alpha_{k+1} + (B_2 - \beta_0) \alpha_k + \gamma_k }{(2k+9)(2k+6)(2k+7)(2k+4)} r^{2k+7}\\
&+ D r + E r^3
\end{split}
\end{equation}
The constants $A$ and $E$ can be determined from the boundary condtions.
The no-stress condition gives
\begin{equation}
  E= \frac{5}{42}A -\frac{115}{24948} +  \sum_{k=0}^{+\infty} \frac{A_2 \alpha_{k+1} + (B_2 - \beta_0) \alpha_k + \gamma_k }{(2k+9)(2k+4)} \label{no-stress_P}
\end{equation}
and continuity of the normal stress gives
\begin{equation}
\begin{split}
&\frac{23}{7}A =- 6 E + \frac{316}{1173}  - \\
&\frac{30}{5} \sum_{k=0}^{+\infty} \frac{A_2 \alpha_{k+1} + (B_2 - \beta_0)\alpha_k + \gamma_k }{(2k+9)(2k+6)(2k+7)(2k+4)}2(k+3)(4k^2+24k+29). \label{normal-stress_P}
\end{split}
\end{equation}
Using equations \eqref{no-stress_P} and  \eqref{normal-stress_P}, we obtain
\begin{equation}
A =  \frac{131}{1764}  -\frac{3}{2} \sum_{k=0}^{+\infty} \frac{A_2 \alpha_{k+1} + (B_2 - \beta_0) \alpha_k + \gamma_k }{2k+7} \simeq 0.0617.
\end{equation}

The average velocity $\bar u_x $  in the $x$ direction, defined as
\begin{equation}
\bar u_x = \frac{1}{V_\mathrm{ic}} \int_{V_\mathrm{ic}} u_x dV
\end{equation}
is less than the infinite viscosity limit (here $V_\mathrm{ic}$ is the volume of the inner core).
Indeed, noting that $u_x = u_r \cos \theta - u_\theta \sin \theta$, and that
\begin{align}
u_r &= \sum_{l,m} l(l+1)\frac{p_l^m}{r}P_l, \\
u_\theta &= \sum_{l,m} \frac{1}{r}\frac{d}{dr}\left( r p_l^m\right) \frac{\partial P_l}{\partial \theta},
\end{align}
we find
\begin{align}
\bar u_x &= \frac{3}{4\pi} \int_0^1\! \int_0^\pi \left[ 2\, p_1 \cos^2 \theta + \frac{d}{dr}\left( r\, p_1\right)  \sin^2 \theta  \right] r \sin\theta dr d\theta d\phi  \\
		&=  \int_0^1 \left[ 4 p_1 +  {2} r \frac{d p_1}{dr} \right] r \,    dr = \int_0^1 2 \frac{d}{dr}\left( r^2 p_1  \right) \,    dr \\
		&   = 2\, p_1(r=1)  \label{TranslationVelocityPoloidal} \\
		& = V_0 \left[ 1 +  \left( A + B + C \right)\mathcal{P} + \mathcal{O}(\mathcal{P}^2) \right] ,
\end{align}
which gives
\begin{equation}
\bar u_x \simeq  \sqrt{\frac{6}{5} \frac{Ra}{{\cal{P}}} } \left[ 1 -0.0216\, {\cal{P}} + \mathcal{O}(\mathcal{P}^2) \right].
\end{equation}


\begin{thebibliography}{}
\expandafter\ifx\csname natexlab\endcsname\relax\def\natexlab#1{#1}\fi

\bibitem[Abramovich \& Stegun(1965)]{abramovich1965}
Abramovich, M. \& Stegun, I., 1965.
\newblock Handbook of mathematical functions, {\it Fourth Printing. Applied
  Math. Ser. 55, US Government Printing Office, Washington DC\/}.

\bibitem[Alboussi\`ere et~al.(2010)Alboussi\`ere, Deguen, \&
  Melzani]{Alboussiere2010}
Alboussi\`ere, T., Deguen, R., \& Melzani, M., 2010.
\newblock Melting induced stratification above the {Earth's} inner core due to
  convective translation, {\it Nature\/}, {\bf 466}, 744--747.

\bibitem[Alboussi\`ere \& Ricard(2013)Alboussi\`ere \&
  Ricard]{Alboussiere2013}
Alboussi\`ere, T. \& Ricard, Y., 2013.
\newblock Reflections on dissipation associated with thermal convection, {\it J. Fluid Mech.\/}, {\bf 725}, doi:10.1017/jfm.2013.241.
  
\bibitem[{Alf{\`e}} et~al.(2002){Alf{\`e}}, {Gillan}, \& {Price}]{Alfe2002}
{Alf{\`e}}, D., {Gillan}, M.~J., \& {Price}, G.~D., 2002.
\newblock {Ab initio chemical potentials of solid and liquid solutions and the
  chemistry of the Earth's core}, {\it J. Chem. Phys.\/}, {\bf 116},
  7127--7136.
  
 \bibitem[{Anderson} \& {Duba}(1997)]{Anderson1997}
Anderson, O.~L. \& {Wenk}, Duba, A., 1997.
\newblock {Experimental melting curve of iron revisited}, {\it
  J. Geophys. Res.\/}, {\bf 102}, 22659--22670.
 
\bibitem[Anufriev et~al.(2005)Anufriev, Jones, \& Soward]{ajs05}
Anufriev, A., Jones, C., \& Soward, A., 2005.
\newblock The {B}oussinesq and anelastic liquid approximations for convection
  in the {E}arth's core, {\it Phys. Earth Planet. Inter.\/},
  {\bf 12}(3), 163--190.
  
 \bibitem[Aubert (2013)Aubert]{Aubert2013}
Aubert, J., 2013.
\newblock  {Flow throughout the Earth's core inverted from geomagnetic observations and numerical dynamo models}, {\it Geophys. J. Int.\/},
  {\bf 192}(2), 537--556.

\bibitem[Bergman et~al.(2010)Bergman, Lewis, Myint, Slivka, Karato, \&
  Abreu]{bergman2010}
Bergman, M., Lewis, D., Myint, I., Slivka, L., Karato, S., \& Abreu, A., 2010.
\newblock {Grain growth and loss of texture during annealing of alloys, and the
  translation of Earth's inner core}, {\it Geophys. Res. Lett.\/},
  {\bf 37}(22), L22313.

\bibitem[Buffett(2009)]{Buffett2009}
Buffett, B., 2009.
\newblock Onset and orientation of convection in the inner core, {\it Geophys.
  J. Int.\/}, {\bf 179}, 711--719.

\bibitem[{Buffett}(1997)]{Buffett1997}
{Buffett}, B.~A., 1997.
\newblock {Geodynamics estimates of the viscosity of the Earth's inner core},
  {\it Nature\/}, {\bf 388}, 571--573.

\bibitem[{Buffett} \& {Wenk}(2001)]{Buffett2001}
{Buffett}, B.~A. \& {Wenk}, H.-R., 2001.
\newblock {Texturing of the Earth's inner core by Maxwell stresses}, {\it
  Nature\/}, {\bf 413}, 60--63.

\bibitem[{Chandrasekhar}(1961)]{Chandrasekhar1961}
{Chandrasekhar}, S., 1961.
\newblock {\it Hydrodynamic and hydromagnetic stability\/}, International
  Series of Monographs on Physics, Oxford: Clarendon.

\bibitem[Christensen \& Aubert(2006)]{Christensen2006}
Christensen, U. \& Aubert, J., 2006.
\newblock Scaling properties of convection-driven dynamos in rotating spherical
  shells and application to planetary magnetic fields, {\it Geophys. J. Int.\/}, {\bf 166}(1), 97--114.

\bibitem[Cottaar \& {Buffett}(2012)]{Cottaar2012}
Cottaar, S. \& {Buffett}, B., 2012.
\newblock Convection in the Earth's inner core, {\it Phys. Earth Planet.
  Inter.\/}, {\bf 198-199}, 67--78.
  
  \bibitem[Davies et~al.(2013)]{Davies2013}
Davies, C.J., Silva, L., \& Mound, J., 2013.
\newblock {On the influence of a translating inner core in models of outer core convection}, {\it Phys. Earth Planet.
  Inter.\/}, {\bf 214}, 104 - 114.
 
\bibitem[{de Koker} et~al.(2012){de Koker}, {Steinle-Neumann}, \&
  Vlcek]{deKoker2012}
{de Koker}, N., {Steinle-Neumann}, G., \& Vlcek, V., 2012.
\newblock Electrical resistivity and thermal conductivity of liquid fe alloys
  at high p and t, and heat flux in earth's core, {\it Proceedings of the
  National Academy of Science\/}, {\bf 109}(11), 4070--4073.

\bibitem[Deguen \& Cardin(2011)]{Deguen2011a}
Deguen, R. \& Cardin, P., 2011.
\newblock {Thermo-chemical convection in Earth's inner core}, {\it Geophys. J.
  Int.\/}, {\bf 187}, 1101--1118.

\bibitem[Deguen et~al.(2011)Deguen, Cardin, Merkel, \& Lebensohn]{Deguen2011b}
Deguen, R., Cardin, P., Merkel, S., \& Lebensohn, R., 2011.
\newblock Texturing in earth's inner core due to preferential growth in its
  equatorial belt, {\it Phys. Earth Planet. Inter.\/}, {\bf 188}, 173--184.

\bibitem[Deschamps et~al.(2012)Deschamps, Yao, Tackley, \&
  Sanchez-Valle]{Deschamps2012}
Deschamps, F., Yao, C., Tackley, P.~J., \& Sanchez-Valle, C., 2012.
\newblock High rayleigh number thermal convection in volumetrically heated
  spherical shells, {\it J. Geophys. Res. - Solid Earth\/}, {\bf
  117}, E09006.

\bibitem[{Dziewonski} \& {Anderson}(1981)]{PREM}
{Dziewonski}, A.~M. \& {Anderson}, D.~L., 1981.
\newblock {Preliminary reference Earth model}, {\it Phys. Earth Planet.
  Inter.\/}, {\bf 25}, 297--356.

\bibitem[Forte \& Peltier(1987)]{FP87}
Forte, A.~M. \& Peltier, W.~R., 1987.
\newblock {Plate Tectonics and Aspherical Earth Structure: The Importance of
  Poloidal-Toroidal Coupling}, {\it J. Geophys. Res.\/}, {\bf
  92}, 3645--3679.

\bibitem[{Geballe} et~al.(2013)]{Geballe2013}
Geballe, Z. M., Lasbleis, M.,  {Cormier}, V.~F. \& Day, E. A. , 2013.
\newblock {Sharp hemisphere boundaries in a translating inner core}, {\it Geophys. Res.
  Lett.\/}, in press.
  
  \bibitem[{Gillet} et~al.(2009){Gillet}, {Pais}, \& {Jault}]{Gillet09}
{Gillet}, N., {Pais}, M.~A. \& {Jault}, D., 2009.
\newblock {Ensemble inversion of time-dependent core flow models}, {\it Geophys. J. Int.\/}, {\bf
  10}, Q06004.

\bibitem[{Gubbins} et~al.(2008){Gubbins}, {Masters}, \& {Nimmo}]{Gubbins08}
{Gubbins}, D., {Masters}, G., \& {Nimmo}, F., 2008.
\newblock {A thermochemical boundary layer at the base of Earth's outer core
  and independent estimate of core heat flux}, {\it Geophys. J. Int.\/}, {\bf
  174}, 1007--1018.

\bibitem[Gubbins et~al.(2011)Gubbins, Sreenivasan, Mound, \& Rost]{Gubbins2011}
Gubbins, D., Sreenivasan, B., Mound, J., \& Rost, S., 2011.
\newblock Melting of the earth's inner core, {\it Nature\/}, {\bf 473}(7347),
  361--363.
  
  \bibitem[Gubbins et~al.(2013)Gubbins, Alf\`e, \& Davies]{gubbins2013}
Gubbins, D., Alf\`e, D., \& Davies, D., 2013.
\newblock {Compositional Instability of Earth's Solid Inner Core}, {\it Geophys. Res.
  Lett.\/}, {\bf 40}(1--5),
  361--363.

\bibitem[Irving et~al.(2009)Irving, Deuss, \& Woodhouse]{irving2009}
Irving, J., Deuss, A., \& Woodhouse, J., 2009.
\newblock Normal mode coupling due to hemispherical anisotropic structure in
  Earth's inner core, {\it Geophys. J. Int.\/}, {\bf 178}(2),
  962--975.

\bibitem[{Jeanloz} \& {Wenk}(1988)]{Jeanloz1988}
{Jeanloz}, R. \& {Wenk}, H.-R., 1988.
\newblock {Convection and anisotropy of the inner core}, {\it Geophys. Res.
  Lett.\/}, {\bf 15}, 72--75.

\bibitem[{Karato}(1999)]{Karato1999}
{Karato}, S.-I., 1999.
\newblock {Seismic anisotropy of the Earth's inner core resulting from flow
  induced by Maxwell stresses}, {\it Nature\/}, {\bf 402}, 871--873.

\bibitem[Koot \& Dumberry(2011)]{Koot2011}
Koot, L. \& Dumberry, M., 2011.
\newblock Viscosity of the earth's inner core: Constraints from nutation
  observations, {\it Earth Planet. Sci. Lett.\/}.

\bibitem[Mizzon \& Monnereau(2013)]{Mizzon2013}
Mizzon, H. \& Monnereau, M., 2013.
\newblock Implication of the lopsided growth for the viscosity of earthÕs inner core, {\it Earth Planet. Sci. Lett.\/}, {\bf 361}, 391--401.

\bibitem[Monnereau et~al.(2010)Monnereau, Calvet, Margerin, \&
  Souriau]{Monnereau2010}
Monnereau, M., Calvet, M., Margerin, L., \& Souriau, A., 2010.
\newblock Lopsided growth of Earth's inner core, {\it Science\/}, {\bf 328},
  1014--1017.

\bibitem[Mound \& Buffett(2006)]{Mound2006}
Mound, J.~E. \& Buffett, B.~A., 2006.
\newblock Detection of a gravitational oscillation in length-of-day, {\it Earth
  Planet. Sci. Lett.\/}, {\bf 243}, 383--389.

\bibitem[Niu \& Wen(2001)]{Niu01}
Niu, F.~L. \& Wen, L.~X., 2001.
\newblock {Hemispherical variations in seismic velocity at the top of the
  Earth's inner core.}, {\it Nature\/}, {\bf 410}, 1081--1084.

\bibitem[Pais et~al.(2008) Pais, \& Jault]{pais2008}
Pais, A. \& Jault, D., 2008.
\newblock Quasi-geostrophic flows responsible for the secular variation of the Earth's magnetic field, {\it Geophys. J. Int.\/}, {\bf 173},
  421--443.

\bibitem[{Parmentier} \& {Sotin}(2000)]{Parmentier2000}
{Parmentier}, E.~M. \& {Sotin}, C., 2000.
\newblock {Three-dimensional numerical experiments on thermal convection in a
  very viscous fluid: Implications for the dynamics of a thermal boundary layer
  at high Rayleigh number}, {\it Physics of Fluids\/}, {\bf 12}, 609--617.

\bibitem[{Poirier}(1994)]{Poirier1994a}
{Poirier}, J.-P., 1994.
\newblock {Physical properties of the Earth's core}, {\it C.R. Acad. Sci.
  Paris\/}, {\bf 318}, 341--350.

\bibitem[Pozzo et~al.(2012)Pozzo, Davies, Gubbins, \& Alf{\`e}]{pozzo2012}
Pozzo, M., Davies, C., Gubbins, D., \& Alf{\`e}, D., 2012.
\newblock Thermal and electrical conductivity of iron at Earth's core
  conditions, {\it Nature\/}, {\bf 485}, 355--358.

\bibitem[Reaman et~al.(2012)Reaman, Colijn, Yang, Hauser, \&
  Panero]{Reaman2012}
Reaman, D., Colijn, H., Yang, F., Hauser, A., \& Panero, W., 2012.
\newblock Interdiffusion of Earth's core materials to 65Gpa and 2200K, {\it
   Earth  Planet. Sci. Lett.\/}, {\bf 349}, 8--14.

\bibitem[Reaman et~al.(2011)Reaman, Daehn, \& Panero]{Reaman2011}
Reaman, D.~M., Daehn, G.~S., \& Panero, W.~R., 2011.
\newblock Predictive mechanism for anisotropy development in the earth's inner
  core, {\it  Earth Planet. Sci. Lett.\/}, {\bf 312}(3-4), 437 --  442.

\bibitem[Ribe(2007)]{Ribe07}
Ribe, N.~M., 2007.
\newblock {\it Analytical Approaches to Mantle Dynamics\/}, In Treatise on
  Geophysics, G. Schubert, Ed., Vol. 7.

\bibitem[Schubert et~al.(2001)Schubert, Turcotte, \& Olson]{sto01}
Schubert, G., Turcotte, D., \& Olson, P., 2001.
\newblock {\it Mantle convection in the Earth and planets\/}, Cambridge
  University Press.

\bibitem[Sha \& Cohen(2011)]{sha2011}
Sha, X. \& Cohen, R., 2011.
\newblock First-principles studies of electrical resistivity of iron under
  pressure, {\it Journal of Physics: Condensed Matter\/}, {\bf 23}, 075401.

\bibitem[Sneddon(1960)]{sneddon1960}
Sneddon, I., 1960.
\newblock On some infinite series involving the zeros of bessel functions of
  the first kind, in {\em Proceedings of the Glasgow Mathematical
  Association\/}, vol.~4, pp. 144--156, Cambridge Univ Press.

\bibitem[{Souriau} \& {Poupinet}(1991)]{Souriau91}
{Souriau}, A. \& {Poupinet}, G., 1991.
\newblock {The velocity profile at the base of the liquid core from
  PKP(BC+Cdiff) data: An argument in favor of radial inhomogeneity}, {\it
  Geophys. Res. Lett.\/}, {\bf 18}, 2023--2026.

\bibitem[Sreenivasan \& Gubbins(2011)]{sreenivasan2011}
Sreenivasan, B. \& Gubbins, D., 2011.
\newblock On mantle-induced heat flow variations at the inner core boundary,
  {\it Phys. Earth Planet. Inter.\/}.

\bibitem[Stacey \& Loper(2007)]{Stacey2007}
Stacey, F. \& Loper, D., 2007.
\newblock A revised estimate of the conductivity of iron alloy at high pressure
  and implications for the core energy balance, {\it Phys. Earth Planet.
  Inter.\/}, {\bf 161}, 13--18.

\bibitem[{Stacey} \& {Anderson}(2001)]{Stacey2001}
{Stacey}, F.~D. \& {Anderson}, O.~L., 2001.
\newblock {Electrical and thermal conductivities of Fe-Ni-Si alloy under core
  conditions}, {\it Phys. Earth Planet. Inter.\/}, {\bf 124}, 153--162.

\bibitem[{Stacey} \& {Davis}(2008)]{Stacey08}
{Stacey}, F.~D. \& {Davis}, P.~M., 2008.
\newblock {\it Physics of the Earth\/}, Cambridge University Press.

\bibitem[Stevenson(1987)]{stevenson87}
Stevenson, D., 1987.
\newblock Limits on lateral density and velocity variations in the earth's
  outer core, {\it Geophysical Journal of the Royal Astronomical Society\/},
  {\bf 88}(1), 311--319.

\bibitem[Sumita et~al.(1995)Sumita, Yoshida, Hamano, \& Kumazawa]{Sumita1995}
Sumita, I., Yoshida, S., Hamano, Y., \& Kumazawa, M., 1995.
\newblock A model for the structural evolution of the earth's core and its
  relation to the observations, in {\em The Earth's central part : Its
  structure and dynamics\/}, pp. 232--260, ed. Yukutake, T.

\bibitem[Tanaka(2012)]{Tanaka2012}
Tanaka, S., 2012.
\newblock Depth extent of hemispherical inner core from pkp (df) and pkp
  (cdiff) for equatorial paths, {\it Phys. Earth Planet. Inter.\/}.

\bibitem[{Tanaka} \& {Hamaguchi}(1997)]{Tanaka97}
{Tanaka}, S. \& {Hamaguchi}, H., 1997.
\newblock {Degree one heterogeneity and hemispherical variation of anisotropy
  in the inner core from PKP(BC)-PKP(DF) times}, {\it J. Geophys. Res.\/}, {\bf 102}, 2925--2938.

\bibitem[Tritton(1988)]{Tritton}
Tritton, D., 1988.
\newblock {\it Physical fluid dynamics\/}, Oxford, Clarendon Press.

\bibitem[{Van Orman}(2004)]{VanOrman04}
{Van Orman}, J.~A., 2004.
\newblock {On the viscosity and creep mechanism of Earth's inner core}, {\it
 Geophys. Res. Lett.\/}, {\bf 31}, 20606--+.

\bibitem[{Vo{\v c}adlo}(2007)]{Vocadlo2007}
{Vo{\v c}adlo}, L., 2007.
\newblock {Ab initio calculations of the elasticity of iron and iron alloys at
  inner core conditions: Evidence for a partially molten inner core?}, {\it
  Earth Planet. Sci. Lett.\/}, {\bf 254}, 227--232.

\bibitem[{Weber} \& {Machetel}(1992)]{Weber92}
{Weber}, P. \& {Machetel}, P., 1992.
\newblock {Convection within the inner-core and thermal implications}, {\it
  Geophys. Res. Lett.\/}, {\bf 19}, 2107--2110.

\bibitem[{Yoshida} et~al.(1996){Yoshida}, {Sumita}, \& {Kumazawa}]{Yoshida1996}
{Yoshida}, S., {Sumita}, I., \& {Kumazawa}, M., 1996.
\newblock {Growth model of the inner core coupled with the outer core dynamics
  and the resulting elastic anisotropy}, {\it J. Geophys. Res.\/}, {\bf 101},
  28085--28104.

\end{thebibliography}
\end{document}